\documentclass[
aps,
prd,
preprint,
superscriptaddress,
nofootinbib,
longbibliography
]{revtex4-2}

\usepackage{amsmath}
\usepackage{amssymb}
\usepackage{bm}
\usepackage{mathtools}
\usepackage{graphicx}
\usepackage{xcolor}
\usepackage{microtype}
\usepackage{hyperref}

\hypersetup{
  colorlinks=true,
  linkcolor=blue,
  citecolor=blue,
  urlcolor=blue,
  pdftitle={Exact constraints on family-separated seesaw relations and their phenomenological consequences},
  pdfauthor={Jianlong Lu},
  pdfsubject={Canonical type-I seesaw and family-separated seesaw relations},
  pdfkeywords={neutrino masses, type-I seesaw, family-separated seesaw, leptogenesis, form dominance}
}

\begin{document}

\title{Exact constraints on family-separated seesaw relations and
their phenomenological consequences}

\author{Jianlong Lu}
\email{jianlong@nus.edu.sg}
\affiliation{
Department of Mathematics, National University of Singapore,
Singapore 119076
}

\date{\today}

\begin{abstract}
Working at tree level and at a common renormalization scale, we examine
the family-separated seesaw ansatz of Z.-z.~Xing,
arXiv:2605.27049v2. For a fixed light--heavy pairing, imposing
$m_iU_{\alpha i}U_{\beta i}+M_iR_{\alpha i}R_{\beta i}=0$ together
with $UU^\dagger+RR^\dagger=\mathbf{1}$ yields the general exact
solution $U=VD^{-1/2}$ and
$R=iVE\,\operatorname{diag}\!\left(\sqrt{r_i/(1+r_i)}\right)$,
where $r_i=m_i/M_i$, $D=\operatorname{diag}(1+r_i)$, $V$ is unitary,
and $E=\operatorname{diag}(\eta_i)$ with $\eta_i=\pm1$. Consequently,
$U^\dagger U$ and $R^\dagger R$ are diagonal, and distinct columns
are exactly orthogonal. An exact unitary completion gives, in a
convenient sterile basis, $M_R=D_N-D_\nu$ and
$Y_\nu=(i/v)VE(D_\nu D_N)^{1/2}$, so that
$Y_\nu^\dagger Y_\nu=D_\nu D_N/v^2$ is diagonal. In the canonical
domain $M_i>m_i$, the singlet masses relevant to the unbroken-phase
decay description are $\widehat{M}_i=M_i-m_i>0$, distinct from the
broken-phase heavy eigenvalues $M_i$. Hence, at the scale of exact
alignment, all standard unflavored and flavored nonresonant one-loop
decay asymmetries vanish in the nondegenerate perturbative regime.
For $m_i=0$, the paired Yukawa column and decay width vanish instead.
The flavor-summed rephasing-invariant combination used in the proposed
asymmetry also cancels identically, although individual CP-odd
invariants may remain nonzero. Thus the proposed correlation between
low-energy CP violation and these decay asymmetries does not follow
from exact family separation. This result does not address radiative
misalignment, resonant or coherent dynamics, or thermal and
higher-loop sources. We also correct the proposed heavy-mass
reconstruction formulas and clarify the implications for
neutrinoless double-beta decay, Yukawa scaling, parameter counting,
and light--heavy mass orderings.
\end{abstract}


\maketitle

\clearpage

\section{Introduction}
\label{sec:introduction}

The canonical type-I seesaw mechanism provides one of the simplest
ultraviolet completions of the dimension-five Weinberg operator and
offers a natural framework for generating Majorana masses for the
observed neutrinos
\cite{Minkowski1977,Yanagida1979,GellMannRamondSlansky1979,
MohapatraSenjanovic1980,Weinberg1979}.
In the three-singlet realization considered here, the neutral-lepton
mass matrix is diagonalized by a $6\times6$ unitary transformation.
Denoting by $U$ and $R$ the $3\times3$ upper submatrices governing the
charged-current interactions of the light and heavy Majorana
neutrinos, respectively, the exact diagonalization implies
\begin{equation}
UD_\nu U^T+RD_NR^T=0,
\qquad
UU^\dagger+RR^\dagger=\mathbf{1},
\label{eq:exact-identities-intro}
\end{equation}
where
\begin{equation}
D_\nu=\operatorname{diag}(m_1,m_2,m_3),
\qquad
D_N=\operatorname{diag}(M_1,M_2,M_3).
\label{eq:mass-matrices-intro}
\end{equation}
Here $m_i$ and $M_i$ denote, respectively, the exact nonnegative
light-neutrino and positive heavy-neutrino Takagi mass eigenvalues of
the complete tree-level post-electroweak-symmetry-breaking
neutral-lepton mass matrix. The first identity in
Eq.~\eqref{eq:exact-identities-intro} follows from the vanishing
active--active block of the canonical type-I neutral-lepton mass
matrix, while the second is the upper-left block of the unitarity
condition for the complete light--heavy mixing matrix
\cite{SchechterValle1980,GrimusLavoura2000,
BlennowFernandezMartinez2011}.
In general, neither $U$ nor $R$ is unitary by itself, and their
entries are nonlinearly correlated through the underlying seesaw
parameters. Throughout this work, ``exact'' refers to algebraic
identities of the renormalized tree-level mass matrix, with all
masses, mixing matrices, and couplings evaluated at a common
renormalization scale.

Throughout this paper, Ref.~\cite{XingFSS} denotes version 2 of
arXiv:2605.27049. That reference proposed a particularly restrictive
solution of the exact seesaw relation, referred to as the
family-separated seesaw ansatz. It requires the contribution of each
paired light and heavy mass eigenstate to vanish separately:
\begin{equation}
m_iU_{\alpha i}U_{\beta i}
+
M_iR_{\alpha i}R_{\beta i}
=
0,
\qquad
i=1,2,3,
\qquad
\alpha,\beta=e,\mu,\tau.
\label{eq:FSS-intro}
\end{equation}
This condition is stronger than the summed matrix relation in
Eq.~\eqref{eq:exact-identities-intro}, which requires only the sum
over the mass-eigenstate index $i$ to vanish. The FSS ansatz was used
in Ref.~\cite{XingFSS} to derive relations between the light- and
heavy-neutrino masses and mixing parameters and, in particular, to
argue for a nonzero connection between low-energy leptonic CP
violation and the CP asymmetries generated in heavy Majorana neutrino
decays.

Equation~\eqref{eq:FSS-intro}, however, aligns the $i$th column of
$R$ with the $i$th column of $U$. Column-aligned seesaw structures
are closely related to the form-dominance condition studied in flavor
models and in the Casas--Ibarra parametrization
\cite{CasasIbarra2001,King2009,ChoubeyKingMitra2010}.
When the corresponding Yukawa-coupling columns are exactly
orthogonal, both the unflavored and flavored standard nonresonant
one-loop decay asymmetries vanish
\cite{ChoubeyKingMitra2010,AristizabalSierra2011}.
It is therefore necessary to impose the FSS condition and exact
partial unitarity simultaneously and to reconstruct the associated
Yukawa structure before inferring singlet-neutrino decay asymmetries
from the active--heavy mixing matrix.

In this work, we derive the complete structure of the upper light-
and heavy-neutrino charged-current mixing blocks implied by
Eqs.~\eqref{eq:exact-identities-intro} and
\eqref{eq:FSS-intro}. Defining the nonnegative ratios
$r_i=m_i/M_i$, we show that the general exact solution for these
upper blocks can be expressed in terms of a single unitary matrix
$V$ as $U=VD^{-1/2}$, where
$D=\operatorname{diag}(1+r_1,1+r_2,1+r_3)$, and
\begin{equation}
R
=
iV\operatorname{diag}\!\left(
\eta_1\sqrt{\frac{r_1}{1+r_1}},
\eta_2\sqrt{\frac{r_2}{1+r_2}},
\eta_3\sqrt{\frac{r_3}{1+r_3}}
\right),
\qquad
\eta_i=\pm1.
\label{eq:Rexact-intro}
\end{equation}
It follows exactly that both $U^\dagger U$ and $R^\dagger R$ are
diagonal and that columns associated with distinct mass eigenstates
are mutually orthogonal.

We further construct the most general unitary completion of the fixed
upper blocks and a fixed light--heavy pairing, up to an arbitrary
unitary choice of sterile weak basis. In a convenient sterile basis in
which the right-handed Majorana mass matrix is diagonal, defining
$E=\operatorname{diag}(\eta_1,\eta_2,\eta_3)$, the exact
reconstruction gives
$M_D=iVE(D_\nu D_N)^{1/2}$ and
$M_{\mathrm R}=D_N-D_\nu$. We henceforth restrict the physical
singlet-neutrino interpretation to the canonical-seesaw domain
$M_i>m_i$, including the intended high-scale regime $M_i\gg v$, so
that the corresponding diagonal right-handed Majorana masses are
positive:
\begin{equation*}
\widehat{M}_i=M_i-m_i>0.
\end{equation*}
We denote the associated Majorana singlet-neutrino mass eigenstates
of $M_{\mathrm R}$ by $\mathcal{N}_i$. The states
$\mathcal{N}_i$ and their masses $\widehat{M}_i$ are the quantities
relevant to the conventional unbroken-phase leptogenesis description,
whereas $N_i$ and $M_i$ denote the exact heavy eigenstates and
positive mass eigenvalues of the complete broken-phase neutral-lepton
mass matrix. In the conventional seesaw hierarchy,
$\widehat{M}_i\simeq M_i$, and $\mathcal{N}_i$ is continuously
connected to the predominantly sterile exact eigenstate $N_i$.

The algebraic relation $M_{\mathrm R}=D_N-D_\nu$ may be formally
continued outside the domain $M_i>m_i$. If $M_i<m_i$, however,
$M_i-m_i$ is a signed diagonal mass parameter rather than a positive
physical Majorana mass, and the corresponding Takagi mass is
$\lvert M_i-m_i\rvert$ after an appropriate field rephasing. If
$M_i=m_i$, the corresponding singular value of $M_{\mathrm R}$
vanishes, and the standard heavy-singlet decay interpretation used
below does not apply.

With the convention $M_D=vY_\nu$, the exact reconstruction implies
\begin{equation}
Y_\nu^\dagger Y_\nu
=
\frac{1}{v^2}D_\nu D_N.
\label{eq:YdagY-exact-intro}
\end{equation}
The Yukawa columns associated with distinct paired indices are
therefore exactly orthogonal, rather than only orthogonal at leading
order in the conventional seesaw expansion. The reconstructed mass
matrix is correspondingly equivalent, at the level of the neutrino
mass sector, to three decoupled one-generation seesaw systems, with
the common unitary matrix $V$ determining their charged-current
flavor directions. In the conventional seesaw limit
$U\simeq V$ and $M_i\simeq\widehat{M}_i$, this structure corresponds,
up to sign conventions and permutations, to a diagonal real
Casas--Ibarra orthogonal matrix, $\mathbb{O}=E$, and hence to form
dominance.

This distinction between the broken- and unbroken-phase quantities is
essential for the decay-asymmetry analysis. Equation~(12) of
Ref.~\cite{XingFSS} describes processes above electroweak symmetry
breaking while labeling the decaying particles by the exact
broken-phase quantities $N_i$ and $M_i$ and expressing their
interactions through the active--heavy mixing matrix $R$. In the
unbroken phase, the relevant decaying states are instead
$\mathcal{N}_i$, their masses are $\widehat{M}_i$, and their decay
amplitudes are governed directly by $Y_\nu$. The two descriptions
coincide only at leading order in the conventional seesaw hierarchy;
the exact relation between $Y_\nu$ and $R$ must be obtained from the
full unitary completion.

This exact Yukawa-column orthogonality has an immediate consequence
for the leptogenesis expressions considered in
Ref.~\cite{XingFSS}. For nondegenerate singlet-neutrino masses
$\widehat{M}_i$ sufficiently far from the resonant regime, all
off-diagonal combinations $(Y_\nu^\dagger Y_\nu)_{ik}$ with
$i\neq k$ vanish exactly, and no standard unflavored or flavored
nonresonant one-loop singlet-neutrino decay asymmetry is generated at
the scale at which exact FSS alignment holds. For a state with a
nonzero Yukawa decay width, the corresponding normalized decay
asymmetries vanish. If $m_i=0$, the paired Yukawa column and its decay
width vanish instead, so that the state generates no decay-produced
lepton asymmetry, although its normalized decay asymmetry is then
undefined. Exact degeneracies among the singlet-neutrino masses
$\widehat{M}_i$ require a separate basis-invariant resonant treatment
and are not described by the standard nonresonant formulas considered
here
\cite{PilaftsisUnderwood2004,
DevMillingtonPilaftsisTeresi2014}.

The preceding vanishing result is a statement about the standard
one-loop decay source evaluated at a common scale where exact FSS
alignment holds. Renormalization-group evolution and threshold
corrections between an alignment scale and the singlet-decay scale
can perturb the Yukawa-column orthogonality and thereby generate
nonzero asymmetries
\cite{CooperKingLuhn2012}. Likewise, the result does not exclude
effects associated with resonant or coherent transport, genuinely
thermal dynamics, or higher-loop sources, which require treatments
beyond the standard nonresonant one-loop decay formulas considered
in Ref.~\cite{XingFSS}. A recent preprint provides an example of a
thermal two-loop source based on lepton-flavor coherences
\cite{LiPilaftsis2026}.

We also show directly that the particular flavor-summed combination
of low-energy rephasing invariants entering the decay-asymmetry
formula of Ref.~\cite{XingFSS} vanishes identically, even though
individual CP-odd invariants may remain nonzero. The proposed nonzero
correlation between low-energy CP violation and standard nonresonant
one-loop singlet-neutrino decay asymmetries therefore does not follow
from the exact FSS structure at the scale at which the alignment
holds.

We also rederive the reconstruction formulas relating the physical
heavy-neutrino mass eigenvalues to the beta-decay mass parameter, the
measured light-neutrino mass-squared differences, and the
active--heavy mixing angles. We identify algebraic errors in all
three heavy-mass formulas presented in Ref.~\cite{XingFSS} and
provide corrected expressions that satisfy the defining beta-decay
relation identically. We also state explicitly the distinction, in
the nonunitary light sector, between the unnormalized beta-decay
coefficient and a normalized weighted endpoint-shape parameter under
the unresolved-spectrum convention. In addition, we clarify the
distinction between the exact seesaw mass-matrix sum rule and the
propagator-dependent neutrinoless double-beta-decay amplitude, the
different heavy-mass scaling of active--heavy mixing and neutrino
Yukawa couplings, the reduction of the physical phase freedom caused
by exact column alignment, and the extent to which the
light-neutrino mass ordering can constrain either the physical heavy
eigenvalues $M_i$ or the corresponding right-handed Majorana
parameters $\widehat{M}_i$.

The remainder of this paper is organized as follows. In
Sec.~\ref{sec:exact-structure}, we derive the general exact FSS
solution for the upper charged-current mixing blocks, construct its
most general unitary completion for a fixed light--heavy pairing, and
obtain the associated exact Yukawa structure and three-pair
decomposition. In Sec.~\ref{sec:leptogenesis}, we analyze the standard
nonresonant one-loop singlet-neutrino decay asymmetries through the
exact Yukawa combinations and through the rephasing invariants used
in Ref.~\cite{XingFSS}. In
Sec.~\ref{sec:mass-reconstruction}, we correct the heavy-neutrino
mass-reconstruction formulas. Further phenomenological implications
are discussed in Sec.~\ref{sec:phenomenology}, and our conclusions
are summarized in Sec.~\ref{sec:conclusions}.

\section{Exact structure of the family-separated solution}
\label{sec:exact-structure}

We first derive, at the common renormalization scale specified in
Sec.~\ref{sec:introduction}, the complete structure of the upper
light- and heavy-neutrino charged-current mixing blocks implied by the
family-separated seesaw condition together with exact partial
unitarity, for a fixed choice of Takagi mass bases and of the
light--heavy pairing. For a fixed mass-eigenstate index $i$, define the
corresponding columns of $U$ and $R$ by
\begin{equation}
\boldsymbol{u}_i
\equiv
\begin{pmatrix}
U_{ei}\\
U_{\mu i}\\
U_{\tau i}
\end{pmatrix},
\qquad
\boldsymbol{\rho}_i
\equiv
\begin{pmatrix}
R_{ei}\\
R_{\mu i}\\
R_{\tau i}
\end{pmatrix}.
\label{eq:column-vectors}
\end{equation}
For each fixed $i$, the FSS condition is the cancellation between two
complex symmetric matrices of rank at most one:
\begin{equation}
m_i\boldsymbol{u}_i\boldsymbol{u}_i^T
+
M_i\boldsymbol{\rho}_i\boldsymbol{\rho}_i^T
=
\mathbf{0}_{3\times3}.
\label{eq:FSS-rank-one}
\end{equation}

Consider first a fixed index $i$ for which $m_i>0$ and $M_i>0$. If
either $\boldsymbol{u}_i$ or $\boldsymbol{\rho}_i$ vanishes,
Eq.~\eqref{eq:FSS-rank-one} requires the other column to vanish as
well. If both columns are nonzero, the image of
$\boldsymbol{u}_i\boldsymbol{u}_i^T$ is
$\operatorname{span}\{\boldsymbol{u}_i\}$, and similarly the image of
$\boldsymbol{\rho}_i\boldsymbol{\rho}_i^T$ is
$\operatorname{span}\{\boldsymbol{\rho}_i\}$. Equation~\eqref{eq:FSS-rank-one}
therefore requires these one-dimensional images to coincide, so the
two nonzero columns are proportional.

The possibility
$\boldsymbol{u}_i=\boldsymbol{\rho}_i=\boldsymbol{0}$ is excluded only
after the FSS condition is combined with exact partial unitarity.
Suppose that one paired column pair vanished identically. For each of
the two remaining values $j\neq i$, the FSS condition implies that
the corresponding paired contribution
\[
\boldsymbol{u}_j\boldsymbol{u}_j^\dagger
+
\boldsymbol{\rho}_j\boldsymbol{\rho}_j^\dagger
\]
has rank at most one. Indeed, if $m_j>0$, the pair either vanishes or
its two nonzero columns are proportional, while if $m_j=0$,
Eq.~\eqref{eq:FSS-rank-one} requires
$\boldsymbol{\rho}_j=\boldsymbol{0}$. The sum of the two remaining
paired contributions would therefore have rank at most two and could
not equal the rank-three identity matrix in
$UU^\dagger+RR^\dagger=\mathbf{1}$. Hence no paired column pair can
vanish identically.

For $m_i>0$ and $M_i>0$, the two columns are therefore nonzero and
proportional:
\begin{equation}
\boldsymbol{\rho}_i
=
\lambda_i\boldsymbol{u}_i.
\label{eq:column-proportionality}
\end{equation}
Substituting Eq.~\eqref{eq:column-proportionality} into
Eq.~\eqref{eq:FSS-rank-one} gives
$\lambda_i^2=-m_i/M_i$. Therefore,
\begin{equation}
R_{\alpha i}
=
i\eta_i
\sqrt{\frac{m_i}{M_i}}\,
U_{\alpha i},
\qquad
\eta_i=\pm1,
\label{eq:column-alignment}
\end{equation}
for every $\alpha=e,\mu,\tau$. The single proportionality coefficient
$\lambda_i$ applies to the entire column, so the same sign $\eta_i$
necessarily applies to all of its flavor components. The two choices
$\eta_i=\pm1$ represent the two square-root branches of
$\lambda_i^2=-m_i/M_i$. The column relation and phase condition
written in Ref.~\cite{XingFSS} select the $+i$ branch; the second
branch differs by an allowed Majorana column-sign redefinition.

The limiting case $m_i=0$ is also included in
Eq.~\eqref{eq:column-alignment}. Since $M_i>0$,
Eq.~\eqref{eq:FSS-rank-one} then requires
$\boldsymbol{\rho}_i=\boldsymbol{0}$, whereas
$\boldsymbol{u}_i$ remains nonzero by the rank argument above. In
this limit, the sign $\eta_i$ is irrelevant. Consequently, any
expression obtained by dividing an FSS relation by $m_i$ is
restricted to $m_i>0$. At $m_i=0$, the corresponding column of $R$
vanishes, and formulas containing explicit factors such as
$M_i/m_i$ must be replaced by their undivided forms. Their values
cannot in general be obtained by direct substitution into the divided
expressions, because the latter may contain indeterminate products
involving vanishing mixing elements.

Define the nonnegative mass ratios
\begin{equation}
r_i
\equiv
\frac{m_i}{M_i},
\label{eq:ri-definition}
\end{equation}
and the diagonal matrix
\begin{equation}
C
\equiv
i\operatorname{diag}\!\left(
\eta_1\sqrt{r_1},
\eta_2\sqrt{r_2},
\eta_3\sqrt{r_3}
\right).
\label{eq:C-definition}
\end{equation}
Equation~\eqref{eq:column-alignment} is equivalent to the exact
matrix relation
\begin{equation}
R
=
UC.
\label{eq:RUC}
\end{equation}

Substituting Eq.~\eqref{eq:RUC} into the exact partial-unitarity
condition gives
\begin{equation}
UU^\dagger+RR^\dagger
=
U\left(
\mathbf{1}+CC^\dagger
\right)U^\dagger
=
\mathbf{1}.
\label{eq:partial-unitarity-substitution}
\end{equation}
Since
\begin{equation}
CC^\dagger
=
\operatorname{diag}(r_1,r_2,r_3),
\label{eq:CCdagger}
\end{equation}
it is convenient to define
\begin{equation}
D
\equiv
\mathbf{1}+CC^\dagger
=
\operatorname{diag}(1+r_1,1+r_2,1+r_3).
\label{eq:D-definition}
\end{equation}
The partial-unitarity condition consequently becomes
\begin{equation}
UDU^\dagger
=
\mathbf{1}.
\label{eq:UDU}
\end{equation}

Because $r_i\geq0$, every eigenvalue $1+r_i$ of $D$ is strictly
positive. The matrices $D^{1/2}$ and $D^{-1/2}$ are therefore
well-defined positive diagonal matrices. Define
\begin{equation}
V
\equiv
UD^{1/2}.
\label{eq:V-definition}
\end{equation}
Using Eq.~\eqref{eq:UDU}, one obtains
\begin{equation}
VV^\dagger
=
UDU^\dagger
=
\mathbf{1}.
\label{eq:V-unitarity}
\end{equation}
Since $V$ is a square matrix, Eq.~\eqref{eq:V-unitarity} implies that
$V$ is unitary and hence that $V^\dagger V=\mathbf{1}$ as well.

For the fixed mass bases and light--heavy pairing specified above, the
general exact solution for the upper light-neutrino charged-current
mixing block is therefore
\begin{equation}
U
=
VD^{-1/2}
=
V\operatorname{diag}\!\left(
\frac{1}{\sqrt{1+r_1}},
\frac{1}{\sqrt{1+r_2}},
\frac{1}{\sqrt{1+r_3}}
\right).
\label{eq:Uexact}
\end{equation}
Using Eq.~\eqref{eq:RUC}, the corresponding upper heavy-neutrino
charged-current mixing block is
\begin{equation}
R
=
iV\operatorname{diag}\!\left(
\eta_1\sqrt{\frac{r_1}{1+r_1}},
\eta_2\sqrt{\frac{r_2}{1+r_2}},
\eta_3\sqrt{\frac{r_3}{1+r_3}}
\right).
\label{eq:Rexact}
\end{equation}

Equations~\eqref{eq:Uexact} and \eqref{eq:Rexact} show that $U$ and
$R$ share the same column directions. Corresponding columns differ
only by their normalization and by the discrete factors $i\eta_i$.
The signs $\eta_i$ correspond to allowed column-sign conventions for
the Majorana mass eigenstates. They do not constitute independent
physical parameters, continuous phases, or additional sources of CP
violation.

The column Gram matrices are
\begin{equation}
U^\dagger U
=
D^{-1}
=
\operatorname{diag}\!\left(
\frac{1}{1+r_1},
\frac{1}{1+r_2},
\frac{1}{1+r_3}
\right),
\label{eq:UdagU}
\end{equation}
and
\begin{equation}
R^\dagger R
=
\operatorname{diag}\!\left(
\frac{r_1}{1+r_1},
\frac{r_2}{1+r_2},
\frac{r_3}{1+r_3}
\right).
\label{eq:RdagR}
\end{equation}
Combining Eqs.~\eqref{eq:UdagU} and \eqref{eq:RdagR} also gives the
additional exact FSS identity
\begin{equation}
U^\dagger U+R^\dagger R
=
\mathbf{1}.
\label{eq:column-partial-unitarity}
\end{equation}
Unlike $UU^\dagger+RR^\dagger=\mathbf{1}$, the relation in
Eq.~\eqref{eq:column-partial-unitarity} is not a generic block
consequence of the unitarity of the full $6\times6$ mixing matrix. It
is a special consequence of exact FSS alignment and of the chosen
one-to-one pairing of the light and heavy columns.

The relative cross-Gram matrix is fixed as well:
\begin{equation}
U^\dagger R
=
iE\operatorname{diag}\!\left(
\frac{\sqrt{r_1}}{1+r_1},
\frac{\sqrt{r_2}}{1+r_2},
\frac{\sqrt{r_3}}{1+r_3}
\right),
\label{eq:UdagR-exact}
\end{equation}
where $E=\operatorname{diag}(\eta_1,\eta_2,\eta_3)$. Thus exact FSS
fixes not only the separate column Gram matrices $U^\dagger U$ and
$R^\dagger R$, but also their relative cross-Gram matrix
$U^\dagger R$.

Consequently, columns associated with distinct mass eigenstates are
exactly orthogonal:
\begin{equation}
(U^\dagger U)_{ik}
=
0,
\qquad
(R^\dagger R)_{ik}
=
0,
\qquad
i\neq k.
\label{eq:column-orthogonality}
\end{equation}
This result does not require the entries of $U$ or $R$ to be real.
The common unitary matrix $V$ may contain nontrivial Dirac-type and
Majorana-type phases, but those phases do not spoil the orthogonality
of distinct columns.

As a consistency check, the exact seesaw relation is automatically
satisfied. Using Eq.~\eqref{eq:RUC}, one has
\begin{equation}
UD_\nu U^T+RD_NR^T
=
U\left(
D_\nu+CD_NC^T
\right)U^T.
\label{eq:seesaw-consistency-start}
\end{equation}
Because $r_iM_i=m_i$ and $\eta_i^2=1$,
\begin{equation}
CD_NC^T
=
-\operatorname{diag}(m_1,m_2,m_3)
=
-D_\nu.
\label{eq:CDNCT}
\end{equation}
It follows that
\begin{equation}
UD_\nu U^T+RD_NR^T
=
\mathbf{0}_{3\times3}.
\label{eq:seesaw-consistency}
\end{equation}

The upper-block solution also admits an exact completion to the full
$6\times6$ unitary diagonalization. To avoid confusion with the
matrix $C$ in Eq.~\eqref{eq:C-definition} and with the lower-left
mixing block conventionally denoted by $S$, define
\begin{equation}
E
\equiv
\operatorname{diag}(\eta_1,\eta_2,\eta_3),
\qquad
\mathsf{C}_r
\equiv
\operatorname{diag}\!\left(
\frac{1}{\sqrt{1+r_1}},
\frac{1}{\sqrt{1+r_2}},
\frac{1}{\sqrt{1+r_3}}
\right),
\label{eq:completion-C-definition}
\end{equation}
and
\begin{equation}
\mathsf{S}_r
\equiv
\operatorname{diag}\!\left(
\sqrt{\frac{r_1}{1+r_1}},
\sqrt{\frac{r_2}{1+r_2}},
\sqrt{\frac{r_3}{1+r_3}}
\right).
\label{eq:completion-S-definition}
\end{equation}
These matrices satisfy
\begin{equation}
\mathsf{C}_r^{\,2}+\mathsf{S}_r^{\,2}
=
\mathbf{1},
\qquad
E^2
=
\mathbf{1}.
\label{eq:completion-identities}
\end{equation}
For an arbitrary unitary $3\times3$ matrix $W$, a unitary completion
of the upper blocks may be written as
\begin{align}
\mathcal{U}
&\equiv
\begin{pmatrix}
U & R\\
S & Q
\end{pmatrix}
\nonumber\\
&=
\begin{pmatrix}
V & \mathbf{0}\\
\mathbf{0} & W
\end{pmatrix}
\begin{pmatrix}
\mathsf{C}_r & iE\mathsf{S}_r\\
iE\mathsf{S}_r & \mathsf{C}_r
\end{pmatrix}
\nonumber\\
&=
\begin{pmatrix}
V\mathsf{C}_r & iVE\mathsf{S}_r\\
iWE\mathsf{S}_r & W\mathsf{C}_r
\end{pmatrix}.
\label{eq:full-unitary-completion}
\end{align}
The middle matrix in Eq.~\eqref{eq:full-unitary-completion} is
unitary by Eq.~\eqref{eq:completion-identities}, and its upper blocks
coincide with Eqs.~\eqref{eq:Uexact} and \eqref{eq:Rexact}.

For the fixed upper blocks, column ordering, and light--heavy pairing,
Eq.~\eqref{eq:full-unitary-completion} gives the most general unitary
completion. Indeed, the lower three rows of any unitary completion
form an orthonormal basis of the orthogonal complement of the fixed
upper three rows. Any two orthonormal bases of this
three-dimensional complement are related by left multiplication by a
unitary $3\times3$ matrix. The arbitrary matrix $W$ therefore
parametrizes precisely this freedom and, within the canonical type-I
neutrino sector considered here, corresponds physically to the choice
of sterile weak basis.

Let the complete complex symmetric neutral-lepton mass matrix be
written as
\begin{equation}
\mathcal{M}
=
\begin{pmatrix}
\mathbf{0} & M_D\\
M_D^T & M_{\mathrm R}
\end{pmatrix},
\label{eq:full-mass-matrix}
\end{equation}
with the Takagi diagonalization convention
\begin{equation}
\mathcal{U}^\dagger\mathcal{M}\mathcal{U}^*
=
\begin{pmatrix}
D_\nu & \mathbf{0}\\
\mathbf{0} & D_N
\end{pmatrix},
\label{eq:full-Takagi-diagonalization}
\end{equation}
or equivalently
\begin{equation}
\mathcal{M}
=
\mathcal{U}
\begin{pmatrix}
D_\nu & \mathbf{0}\\
\mathbf{0} & D_N
\end{pmatrix}
\mathcal{U}^T.
\label{eq:full-mass-reconstruction}
\end{equation}
Substituting Eq.~\eqref{eq:full-unitary-completion} into
Eq.~\eqref{eq:full-mass-reconstruction} gives the exact off-diagonal
Dirac mass matrix
\begin{equation}
M_D
=
iVE\left(D_\nu D_N\right)^{1/2}W^T,
\label{eq:MD-exact-FSS}
\end{equation}
and the right-handed Majorana mass matrix
\begin{equation}
M_{\mathrm R}
=
W\left(D_N-D_\nu\right)W^T.
\label{eq:MR-exact-FSS}
\end{equation}
The vanishing upper-left block is precisely the exact seesaw relation
already verified in Eq.~\eqref{eq:seesaw-consistency}.

Because the sterile fields carry no Standard Model gauge charges, one
may use the sterile weak-basis freedom to set $W=\mathbf{1}$. In this
basis,
\begin{equation}
M_D
=
iVE\left(D_\nu D_N\right)^{1/2},
\qquad
M_{\mathrm R}
=
D_N-D_\nu.
\label{eq:MD-MR-diagonal-basis}
\end{equation}
Thus the diagonal right-handed Majorana mass parameters are
\begin{equation}
\widehat{M}_i
=
M_i-m_i.
\label{eq:Mhat-definition}
\end{equation}
In the canonical-seesaw domain $M_i>m_i$ adopted for the physical
singlet-neutrino interpretation, these parameters are positive Takagi
masses. We denote the corresponding Majorana singlet-neutrino mass
eigenstates by $\mathcal{N}_i$. In the usual hierarchy
$m_i\ll M_i$, one has $\widehat{M}_i\simeq M_i$. The masses entering
the standard unbroken-phase singlet-neutrino decay formulas are
therefore $\widehat{M}_i$, whereas $M_i$ are the exact positive heavy
eigenvalues of the complete broken-phase neutral-lepton mass matrix.

The algebraic reconstruction itself remains valid if it is formally
continued outside the domain $M_i>m_i$. If $M_i<m_i$, however,
$\widehat{M}_i=M_i-m_i$ is a signed diagonal mass parameter rather
than a positive physical mass; the corresponding Takagi mass is
$\lvert M_i-m_i\rvert$ after an appropriate sterile-field rephasing.
If $M_i=m_i$, the corresponding singular value of
$M_{\mathrm R}$ vanishes, and the conventional heavy-singlet decay
interpretation does not apply.

The reconstructed mass matrix also makes the structural meaning of
family separation explicit. At the level of the neutrino mass matrix
alone, after the sterile weak-basis choice $W=\mathbf{1}$ and the
complex-symmetric congruence transformation generated in the active
subspace by $V^*$, followed by a permutation that groups each paired
light and heavy index together, the mass matrix becomes
\[
\mathcal{M}
\ \sim\
\bigoplus_{i=1}^{3}
\begin{pmatrix}
0
&
i\eta_i\sqrt{m_iM_i}
\\
i\eta_i\sqrt{m_iM_i}
&
M_i-m_i
\end{pmatrix}.
\]
Each $2\times2$ complex symmetric block has Takagi singular values
$m_i$ and $M_i$. Exact FSS is therefore equivalent, at the level of
the neutrino mass matrix, to three decoupled one-generation seesaw
systems. The common unitary matrix $V$ supplies their charged-current
flavor directions when the charged-lepton mass basis is restored.

Using the convention $M_D=vY_\nu$,
Eq.~\eqref{eq:MD-MR-diagonal-basis} gives
\begin{equation}
Y_\nu
=
\frac{i}{v}
VE\left(D_\nu D_N\right)^{1/2}.
\label{eq:Yukawa-exact-FSS}
\end{equation}
Consequently,
\begin{equation}
Y_\nu^\dagger Y_\nu
=
\frac{1}{v^2}D_\nu D_N
=
\frac{1}{v^2}
\operatorname{diag}
\left(
m_1M_1,
m_2M_2,
m_3M_3
\right).
\label{eq:YdagY-exact-FSS}
\end{equation}
The Yukawa columns associated with distinct paired indices are
therefore exactly orthogonal in the $W=\mathbf{1}$ singlet basis, in
which $M_{\mathrm R}$ and $Y_\nu^\dagger Y_\nu$ are simultaneously
diagonal. For nondegenerate $\widehat{M}_i$, this singlet mass basis
is fixed up to permutations and allowed Majorana sign choices.

In the conventional seesaw limit, $U\simeq V$ and
$M_i\simeq\widehat{M}_i$, so
Eq.~\eqref{eq:Yukawa-exact-FSS} corresponds, up to
convention-dependent signs and permutations, to a diagonal real
Casas--Ibarra orthogonal matrix,
\begin{equation}
\mathbb{O}
=
E.
\label{eq:CI-form-dominance}
\end{equation}
The exact FSS reconstruction is therefore the all-orders
mass-matrix realization of the Yukawa-column orthogonality associated
with form dominance.

Equations~\eqref{eq:Rexact} and \eqref{eq:Yukawa-exact-FSS} also give
the exact relation
\begin{equation}
R
=
vY_\nu
\left[
D_N\left(D_N+D_\nu\right)
\right]^{-1/2},
\label{eq:R-Y-exact}
\end{equation}
or, componentwise,
\begin{equation}
R_{\alpha i}
=
\frac{v(Y_\nu)_{\alpha i}}
{\sqrt{M_i(M_i+m_i)}}.
\label{eq:R-Y-exact-components}
\end{equation}
In the conventional seesaw limit $m_i\ll M_i$, this reduces to the
familiar leading-order relation
$R_{\alpha i}\simeq v(Y_\nu)_{\alpha i}/M_i$.

The derivation above presupposes a definite choice of Takagi mass
bases and of the light--heavy pairing. If $D_\nu$ or $D_N$ contains
an exact degeneracy, real orthogonal rotations within the
corresponding degenerate Majorana subspace preserve the diagonal mass
matrix but can change the individual columns of $U$ or $R$. The FSS
relations remain valid in a chosen paired basis, but an individual
family pairing is not basis invariant until that basis has been
specified. The most-general statements above should therefore be
understood after the mass bases, column ordering, and pairing have
been fixed.

Separately, an exact degeneracy among the positive right-handed
Majorana masses $\widehat{M}_i$ introduces a residual real orthogonal
freedom in the singlet mass basis. Under such a rotation,
$M_{\mathrm R}$ remains diagonal, whereas the matrix representation
of $Y_\nu^\dagger Y_\nu$ need not remain diagonal if the corresponding
values $m_iM_i$ are unequal. The $W=\mathbf{1}$ construction selects
a simultaneous diagonal basis furnished by the exact completion, but
statements about individual states within an exactly degenerate
subspace require a basis-invariant treatment. Since
$\widehat{M}_i=M_i-m_i$, degeneracy among the physical broken-phase
heavy eigenvalues $M_i$ and degeneracy among the singlet masses
$\widehat{M}_i$ need not coincide.

The FSS ansatz together with exact partial unitarity therefore does
not leave the upper light- and heavy-neutrino charged-current mixing
blocks independently adjustable. Up to the common unitary matrix
$V$, their column normalizations are fixed completely by the three
ratios $r_i=m_i/M_i$. Moreover, for fixed mass bases, column ordering,
and light--heavy pairing, the most general unitary completion fixes
the associated Yukawa structure up to sterile weak-basis freedom and
provides a singlet mass basis in which
$Y_\nu^\dagger Y_\nu$ is exactly diagonal. This exact
Yukawa-column orthogonality will be central to the standard one-loop
leptogenesis analysis in the following section.

\section{Vanishing of the standard nonresonant one-loop
singlet-neutrino decay asymmetries}
\label{sec:leptogenesis}

We now examine, at the common renormalization scale specified in
Sec.~\ref{sec:introduction}, the implications of the exact column
orthogonality and the associated exact Yukawa structure derived in
Sec.~\ref{sec:exact-structure} for the standard nonresonant one-loop
decay asymmetries of the Majorana singlet neutrinos. Two logically
distinct statements will be established. First, in the
charged-lepton mass basis and, for nondegenerate singlet masses, the
singlet mass basis selected in Sec.~\ref{sec:exact-structure}, exact
FSS alignment makes $Y_\nu^\dagger Y_\nu$ diagonal. The standard
unflavored and flavored nonresonant one-loop decay asymmetries
therefore vanish exactly for every singlet state with a nonzero
Yukawa decay width. Second, the particular flavor-summed combination
of rephasing invariants appearing in the FSS decay-asymmetry formula
vanishes exactly as a consequence of the orthogonality of the columns
of $U$.

\subsection{Unflavored decay asymmetries}
\label{subsec:unflavoured-asymmetries}

Let $Y_\nu$ denote the neutrino Yukawa coupling matrix in the basis in
which the charged-lepton mass matrix and the right-handed Majorana
mass matrix are diagonal, and define
\begin{equation}
H_\nu
\equiv
Y_\nu^\dagger Y_\nu.
\label{eq:Hnu-definition}
\end{equation}
To distinguish the two mass notions introduced by the exact
completion, we denote by $\mathcal{N}_i$ the Majorana singlet-neutrino
mass eigenstates of $M_{\mathrm R}$ and by
$\widehat{M}_i$ their positive Takagi masses. The symbols $N_i$ and
$M_i$ continue to denote the exact heavy eigenstates and positive
Takagi mass eigenvalues of the complete tree-level
post-electroweak-symmetry-breaking neutral-lepton mass matrix. In the
exact reconstruction obtained in Sec.~\ref{sec:exact-structure},
\begin{equation}
\widehat{M}_i
=
M_i-m_i.
\label{eq:Mhat-leptogenesis}
\end{equation}
Here and throughout this section, we adopt the canonical-seesaw
domain $M_i>m_i$, so that $\widehat{M}_i>0$. The FSS construction
supplies a definite pairing among the light state $\nu_i$, the exact
heavy state $N_i$, and the singlet column associated with
$\mathcal{N}_i$. In the conventional seesaw regime $m_i\ll M_i$, one
has $\widehat{M}_i\simeq M_i$, and $\mathcal{N}_i$ is continuously
connected to the predominantly sterile exact eigenstate $N_i$.
When applying the conventional unbroken-phase decay formulas below,
we additionally assume that the relevant decay epoch lies in the
electroweak-symmetric phase, as in the intended high-scale seesaw
regime. The inequality $M_i>m_i$ by itself ensures a positive
singlet Majorana mass but does not by itself guarantee that the
unbroken-phase thermal description is applicable.

This distinction is essential for interpreting the decay formula in
Ref.~\cite{XingFSS}. Equation~(12) of that reference describes decays
above electroweak symmetry breaking while labeling the decaying
particles by the exact broken-phase quantities $N_i$ and $M_i$ and
expressing their interactions through the active--heavy mixing matrix
$R$. In the unbroken phase, the relevant decaying particles are
instead $\mathcal{N}_i$, their masses are $\widehat{M}_i$, and their
decay amplitudes are governed directly by $Y_\nu$. The two
descriptions agree only at leading order in the conventional seesaw
hierarchy; the exact connection between them follows from the unitary
completion derived in Sec.~\ref{sec:exact-structure}.

For nondegenerate positive singlet-neutrino masses
$\widehat{M}_i$ sufficiently far from the resonant regime, the
standard one-loop unflavored decay asymmetry generated by the decays
of $\mathcal{N}_i$ has the form
\begin{equation}
\varepsilon_i
=
\frac{1}{8\pi (H_\nu)_{ii}}
\sum_{k\neq i}
\operatorname{Im}\!\left[
(H_\nu)_{ik}^{\,2}
\right]
f(x_{ki}),
\label{eq:eps-standard}
\end{equation}
where
\begin{equation}
x_{ki}
\equiv
\frac{\widehat{M}_k^2}{\widehat{M}_i^2},
\label{eq:xki-definition}
\end{equation}
and $f(x_{ki})$ contains the vertex and self-energy loop
contributions
\cite{CoviRouletVissani1996,DavidsonNardiNir2008,
FongNardiRiotto2012}. The detailed form of the loop function is not
needed for the argument below. The nearly degenerate resonant regime
requires a resummed, basis-invariant treatment of unstable-state
mixing and, in general, flavor-coherent transport, and is not
described by Eq.~\eqref{eq:eps-standard}
\cite{PilaftsisUnderwood2004,
DevMillingtonPilaftsisTeresi2014}. The present discussion concerns
the standard perturbative nonresonant one-loop expressions considered
in Ref.~\cite{XingFSS}.

For nondegenerate $\widehat{M}_i$, the basis in which
$M_{\mathrm R}$ is diagonal is fixed up to permutations and allowed
Majorana sign choices, and the $W=\mathbf{1}$ construction of
Sec.~\ref{sec:exact-structure} makes $H_\nu$ exactly diagonal in that
basis. If an exact degeneracy among the $\widehat{M}_i$ is present,
residual real orthogonal rotations within the degenerate Majorana
subspace introduce an additional basis freedom and can change the
matrix representation of $H_\nu$. The $W=\mathbf{1}$ construction
still supplies a simultaneous diagonal basis, but individual states
inside the degenerate subspace are not uniquely defined.
Equation~\eqref{eq:eps-standard} is then inapplicable, and a separate
basis-invariant resonant analysis is required. No conclusion about an
exactly degenerate resonant source is inferred here merely from the
numerator of the nonresonant formula.

The exact relation between the active--heavy mixing matrix and the
neutrino Yukawa matrix derived in
Eq.~\eqref{eq:R-Y-exact-components} is
\begin{equation}
(Y_\nu)_{\alpha i}
=
\frac{\sqrt{M_i(M_i+m_i)}}{v}\,
R_{\alpha i}.
\label{eq:Y-R-exact-leptogenesis}
\end{equation}
It follows exactly that
\begin{align}
(H_\nu)_{ik}
&=
\sum_{\alpha}
(Y_\nu)_{\alpha i}^{*}
(Y_\nu)_{\alpha k}
\nonumber\\
&=
\frac{
\sqrt{M_i(M_i+m_i)M_k(M_k+m_k)}
}{v^2}
\sum_{\alpha}
R_{\alpha i}^{*}R_{\alpha k}
\nonumber\\
&=
\frac{
\sqrt{M_i(M_i+m_i)M_k(M_k+m_k)}
}{v^2}
(R^\dagger R)_{ik}.
\label{eq:H-R}
\end{align}

The exact FSS solution derived in
Sec.~\ref{sec:exact-structure} satisfies
\begin{equation}
(R^\dagger R)_{ik}
=
0,
\qquad
i\neq k.
\label{eq:RdagR-offdiag-zero}
\end{equation}
Consequently,
\begin{equation}
(H_\nu)_{ik}
=
0,
\qquad
i\neq k.
\label{eq:H-offdiag-zero}
\end{equation}
For the diagonal elements, Eqs.~\eqref{eq:H-R} and
\eqref{eq:RdagR} give
\begin{equation}
(H_\nu)_{ii}
=
\frac{M_i(M_i+m_i)}{v^2}
\frac{r_i}{1+r_i}
=
\frac{m_iM_i}{v^2}.
\label{eq:H-diagonal-exact}
\end{equation}
Equation~\eqref{eq:H-diagonal-exact} agrees with the exact matrix
identity
$Y_\nu^\dagger Y_\nu=D_\nu D_N/v^2$ obtained in
Eq.~\eqref{eq:YdagY-exact-FSS}.

Every interference term in Eq.~\eqref{eq:eps-standard} therefore
vanishes exactly. For each nondegenerate singlet state with a nonzero
Yukawa decay width, one obtains
\begin{equation}
\varepsilon_i
=
0,
\qquad
(H_\nu)_{ii}>0.
\label{eq:eps-zero}
\end{equation}
The tree-level Yukawa decay width of $\mathcal{N}_i$ is proportional
to $\widehat{M}_i(H_\nu)_{ii}$. If $m_i=0$, exact FSS instead implies
that the entire $i$th Yukawa column vanishes:
\begin{equation}
(Y_\nu)_{\alpha i}
=
0,
\qquad
(H_\nu)_{ii}
=
0.
\label{eq:massless-Yukawa-column}
\end{equation}
The corresponding singlet state then has no Yukawa decay width and
generates no decay-produced lepton asymmetry. Its normalized quantity
$\varepsilon_i$, which would divide by $(H_\nu)_{ii}$, is undefined
rather than a nonzero or finite decay asymmetry.

This is the familiar consequence of an exactly column-aligned, or
form-dominant, seesaw structure: mutually orthogonal Yukawa columns
eliminate the CP-odd interference required for the standard
nonresonant one-loop decay asymmetry
\cite{King2009,ChoubeyKingMitra2010}. The conclusion does not require
the individual entries of $U$, $R$, or $Y_\nu$ to be real. These
matrices may contain nontrivial CP-violating phases, but those phases
do not produce a nonzero unflavored nonresonant one-loop decay
asymmetry when the relevant Yukawa columns are orthogonal.

Unlike an argument based only on the leading-order approximation
$R_{\alpha i}\simeq v(Y_\nu)_{\alpha i}/M_i$, the conclusion above
follows from the exact unitary completion of the FSS mixing
structure. Algebraic subleading terms in the exact relation between
$Y_\nu$ and $R$ therefore cannot restore a standard nonresonant
one-loop decay asymmetry while the exact FSS relations remain
satisfied at the same scale. This statement concerns the
nondegenerate perturbative regime and does not replace the separate
basis-invariant resonant analysis required for degenerate or
sufficiently quasi-degenerate singlet states.

\subsection{Flavored decay asymmetries}
\label{subsec:flavoured-asymmetries}

When charged-lepton flavor effects are resolved, the decay asymmetry
of $\mathcal{N}_i$ into a definite charged-lepton flavor $\alpha$ has
the schematic nonresonant one-loop structure
\begin{align}
\varepsilon_{i\alpha}
={}&
\frac{1}{8\pi (H_\nu)_{ii}}
\sum_{k\neq i}
\Bigg\{
\operatorname{Im}\!\left[
(Y_\nu)_{\alpha i}^{*}
(Y_\nu)_{\alpha k}
(H_\nu)_{ik}
\right]
f^{(1)}(x_{ki})
\nonumber\\
&\hspace{33mm}
+
\operatorname{Im}\!\left[
(Y_\nu)_{\alpha i}^{*}
(Y_\nu)_{\alpha k}
(H_\nu)_{ki}
\right]
f^{(2)}(x_{ki})
\Bigg\}.
\label{eq:flavoured-asymmetry}
\end{align}
Here $f^{(1)}$ and $f^{(2)}$ denote the relevant loop functions
\cite{Abada2006,NardiNirRouletRacker2006,
DavidsonNardiNir2008}. Equation~\eqref{eq:flavoured-asymmetry} is
written only to display the Yukawa structures relevant to the
cancellation. Overall signs and the precise definitions of the two
loop functions depend on the convention adopted. In every standard
nonresonant convention relevant here, both interference structures
contain an off-diagonal element of $H_\nu$.

Since $H_\nu$ is Hermitian,
\begin{equation}
(H_\nu)_{ki}
=
(H_\nu)_{ik}^{*},
\label{eq:Hnu-hermiticity}
\end{equation}
and Eq.~\eqref{eq:H-offdiag-zero} implies that both off-diagonal
factors vanish exactly. Hence, for every nondegenerate singlet state
with a nonzero Yukawa decay width,
\begin{equation}
\varepsilon_{i\alpha}
=
0,
\qquad
(H_\nu)_{ii}>0,
\qquad
\alpha=e,\mu,\tau.
\label{eq:flavoured-asymmetry-zero}
\end{equation}
If $m_i=0$, the corresponding Yukawa column and decay width vanish as
in Eq.~\eqref{eq:massless-Yukawa-column}; that singlet state therefore
generates no flavored decay asymmetry, although the normalized
quantity $\varepsilon_{i\alpha}$ is then undefined. Resolving the
charged-lepton flavors consequently does not evade the cancellation.
This agrees with the established result that exact form dominance
eliminates both the total and flavor-dependent standard
nonresonant one-loop leptogenesis decay asymmetries
\cite{ChoubeyKingMitra2010}.

\subsection{Exact cancellation in terms of the rephasing invariants}
\label{subsec:invariant-cancellation}

The cancellation can also be demonstrated directly in the
rephasing-invariant notation used in Ref.~\cite{XingFSS}.
Equation~(12) of Ref.~\cite{XingFSS} contains
$\mathcal{V}_{\alpha\beta}^{jk}$ in its second flavor sum, although
the index $j$ is not defined in that expression. We interpret this
superscript as $ik$, which is the index structure obtained directly
by expanding
$\operatorname{Im}[(R^\dagger R)_{ik}^{\,2}]$.

Define
\begin{equation}
\mathcal{V}_{\alpha\beta}^{ik}
\equiv
\operatorname{Im}\!\left(
U_{\alpha i}U_{\beta i}
U_{\alpha k}^{*}U_{\beta k}^{*}
\right).
\label{eq:V-invariant-definition}
\end{equation}
The symbol $\mathcal{V}_{\alpha\beta}^{ik}$ is used here to avoid
confusion with the unitary matrix $V$ introduced in
Eq.~\eqref{eq:Uexact}.

For $i\neq k$, the flavor-summed combination relevant to the
unflavored decay asymmetry satisfies
\begin{equation}
\operatorname{Im}\!\left[
\left(
\sum_{\alpha}
R_{\alpha i}^{*}R_{\alpha k}
\right)^2
\right]
=
\operatorname{Im}\!\left[
(R^\dagger R)_{ik}^{\,2}
\right].
\label{eq:R-invariant-start}
\end{equation}
Using Eqs.~\eqref{eq:column-alignment} and
\eqref{eq:ri-definition}, one finds
\begin{align}
\operatorname{Im}\!\left[
(R^\dagger R)_{ik}^{\,2}
\right]
={}&
r_i r_k\,
\operatorname{Im}\!\left[
\left(
\sum_{\alpha}
U_{\alpha i}^{*}U_{\alpha k}
\right)^2
\right]
\nonumber\\
={}&
-r_i r_k
\left[
\sum_{\alpha}
\mathcal{V}_{\alpha\alpha}^{ik}
+
2\sum_{\alpha<\beta}
\mathcal{V}_{\alpha\beta}^{ik}
\right].
\label{eq:invariant-combination}
\end{align}
The relative minus sign arises because
$U_{\alpha i}^{*}U_{\alpha k}
U_{\beta i}^{*}U_{\beta k}$
is the complex conjugate of
$U_{\alpha i}U_{\beta i}
U_{\alpha k}^{*}U_{\beta k}^{*}$.

The exact FSS solution also gives
\begin{equation}
\sum_{\alpha}
U_{\alpha i}^{*}U_{\alpha k}
=
(U^\dagger U)_{ik}
=
0,
\qquad
i\neq k.
\label{eq:U-column-orthogonality}
\end{equation}
Therefore,
\begin{equation}
\sum_{\alpha}
\mathcal{V}_{\alpha\alpha}^{ik}
+
2\sum_{\alpha<\beta}
\mathcal{V}_{\alpha\beta}^{ik}
=
0,
\qquad
i\neq k.
\label{eq:V-invariant-sum-rule}
\end{equation}

This sum rule is exact at the level of the FSS mixing matrices
evaluated at the common scale at which exact FSS alignment is
imposed. The individual invariants
$\mathcal{V}_{\alpha\beta}^{ik}$ may remain nonzero because the FSS
solution does not require the Dirac-type or Majorana-type phases in
$U$ to vanish. Nevertheless, the specific flavor-summed combination
entering the decay-asymmetry expression cancels identically.
Rewriting the asymmetry in terms of individually nonzero invariants
therefore cannot by itself produce a nonzero result.

In particular, the decay-asymmetry expression presented in
Ref.~\cite{XingFSS} is proportional to the combination in
Eq.~\eqref{eq:V-invariant-sum-rule}. Once exact FSS alignment and
partial unitarity are imposed simultaneously, that combination
vanishes. Thus the proposed nonzero correlation between low-energy
leptonic CP violation and standard nonresonant one-loop
singlet-neutrino decay asymmetries does not follow from the exact FSS
structure at the scale at which the alignment holds. A nonzero
standard nonresonant one-loop asymmetry would require a departure
from the corresponding orthogonal Yukawa-column structure at the
singlet-decay scale or additional interactions contributing to the
decay source
\cite{ChoubeyKingMitra2010}.

The result established in this section is not a general no-go theorem
for every possible leptogenesis mechanism. Renormalization-group
evolution and threshold corrections between an FSS alignment scale
and the singlet-decay scale can perturb the exact column
orthogonality
\cite{CooperKingLuhn2012}. Degenerate or sufficiently
quasi-degenerate singlet states require a resummed, basis-invariant,
flavor-coherent treatment
\cite{PilaftsisUnderwood2004,
DevMillingtonPilaftsisTeresi2014}. Genuinely thermal or higher-loop
sources likewise lie outside the standard nonresonant one-loop decay
formulas analyzed here. A recent preprint provides an example of a
thermal two-loop source based on lepton-flavor coherences
\cite{LiPilaftsis2026}. These qualifications do not affect the
algebraic cancellation of the particular one-loop expression proposed
in Ref.~\cite{XingFSS}.

\section{Correction of the heavy-neutrino mass formulas}
\label{sec:mass-reconstruction}

We next reconsider the reconstruction of the physical heavy Majorana
neutrino mass eigenvalues from the beta-decay mass parameter, the
light-neutrino mass-squared differences, and the active--heavy mixing
angles. Throughout this section, we use the notation
$c_{ij}\equiv\cos\theta_{ij}$,
$s_{ij}\equiv\sin\theta_{ij}$, and
$t_{ij}\equiv\tan\theta_{ij}$. We also define
\begin{equation}
C_e
\equiv
c_{14}c_{15}c_{16}.
\label{eq:Ce-definition}
\end{equation}
Here and below, $M_i$ denotes the exact positive heavy-neutrino
Takagi mass eigenvalue appearing in $D_N$, with the index $i$
referring to the fixed light--heavy pairing used in the FSS ansatz.
It should be distinguished from the corresponding right-handed
Majorana mass parameter
$\widehat{M}_i=M_i-m_i$ defined in
Eq.~\eqref{eq:Mhat-definition}. As in the preceding sections,
``exact'' refers to the algebraic tree-level relations of the
complete broken-phase mass matrix, with the parameters entering a
given relation evaluated consistently.

For the conventional canonical-seesaw interpretation, we restrict
attention to the domain
\begin{equation}
M_i>m_i,
\label{eq:canonical-domain-mass-reconstruction}
\end{equation}
so that $\widehat{M}_i=M_i-m_i>0$ is the positive Takagi mass of the
corresponding singlet state $\mathcal{N}_i$ in the basis where
$M_{\mathrm R}$ is diagonal. If the algebraic solution is formally
continued to $M_i<m_i$, the corresponding diagonal entry
$M_i-m_i$ is negative and its physical Takagi singular value is
$|M_i-m_i|$ after an appropriate field rephasing. At $M_i=m_i$, the
corresponding singlet Majorana mass parameter vanishes. These formal
domains do not have the conventional heavy-singlet interpretation
used in standard seesaw phenomenology.

In the complete Euler-like parametrization adopted in
Ref.~\cite{XingFSS}, the first row of the light-neutrino mixing matrix
has the exact form
\begin{equation}
|U_{e1}|^2
=
C_e^2c_{12}^2c_{13}^2,
\qquad
|U_{e2}|^2
=
C_e^2s_{12}^2c_{13}^2,
\qquad
|U_{e3}|^2
=
C_e^2s_{13}^2.
\label{eq:first-row-moduli}
\end{equation}
Consequently,
\begin{equation}
\sum_{i=1}^{3}|U_{ei}|^2
=
C_e^2.
\label{eq:first-row-norm}
\end{equation}
Thus the light-neutrino part of the electron row is not normalized to
unity when active--heavy mixing is present. Here $\theta_{12}$ and
$\theta_{13}$ are parameters of the adopted complete light--heavy
mixing parametrization. A phenomenological application of the
reconstruction formulas requires values obtained from an analysis in
which possible light-sector nonunitarity is treated consistently,
rather than identifying them without qualification with parameters
extracted under an exactly unitary three-neutrino hypothesis.

Assuming that the three physical heavy neutrinos are kinematically
inaccessible in the beta-decay process under consideration, and
adopting the unnormalized convention used in
Ref.~\cite{XingFSS}, we define
\begin{equation}
m_\beta^2
\equiv
\sum_{i=1}^{3}m_i^2|U_{ei}|^2.
\label{eq:mbeta-definition}
\end{equation}
Possible effects of nonunitary leptonic mixing on direct kinematic
neutrino-mass observables have been discussed in
Ref.~\cite{Rodejohann2010}. For the particular unresolved-spectrum
fitting convention adopted here, in which the overall spectral
normalization is treated independently, it is useful to distinguish
the unnormalized coefficient in Eq.~\eqref{eq:mbeta-definition} from
a normalized weighted endpoint-shape parameter. Provided
$C_e\neq0$, we define
\begin{equation}
m_{\beta,\mathrm{norm}}^2
\equiv
\frac{\displaystyle\sum_{i=1}^{3}m_i^2|U_{ei}|^2}
{\displaystyle\sum_{i=1}^{3}|U_{ei}|^2}
=
\frac{m_\beta^2}{C_e^2}.
\label{eq:normalized-mbeta-definition}
\end{equation}
The origin of this normalized quantity can be seen from the
mass-dependent kinematic factor in an unresolved beta-decay spectrum.
Writing $w$ for the available neutrino energy and expanding in the
regime $w^2\gg m_i^2$, one obtains schematically
\begin{align}
\sum_{i=1}^{3}
|U_{ei}|^2w\sqrt{w^2-m_i^2}
={}&
w^2\sum_{i=1}^{3}|U_{ei}|^2
-
\frac{1}{2}
\sum_{i=1}^{3}m_i^2|U_{ei}|^2
\nonumber\\
&+
\mathcal{O}\!\left(
\frac{\sum_i |U_{ei}|^2m_i^4}{w^2}
\right).
\label{eq:unresolved-spectrum-expansion}
\end{align}
After the overall normalization
$\sum_i|U_{ei}|^2$ is treated independently, the leading relative
mass-dependent shape correction is therefore governed by the ratio
in Eq.~\eqref{eq:normalized-mbeta-definition}.

This normalized quantity characterizes the leading mass-dependent
endpoint-shape correction under the stated unresolved-spectrum
convention. It should not be interpreted as a replacement for a
complete spectral likelihood analysis. If the individual
light-neutrino mass eigenstates are spectrally resolved, their masses
and weights enter separately. Moreover, an experimental limit quoted
under an exactly unitary three-neutrino hypothesis cannot in general
be identified directly with either
Eq.~\eqref{eq:mbeta-definition} or
Eq.~\eqref{eq:normalized-mbeta-definition} without accounting for
the treatment of the spectral normalization and possible
nonunitarity in the experimental analysis. For direct comparison
with Ref.~\cite{XingFSS}, the unnormalized quantity $m_\beta^2$ is
retained in the definitions. The reconstruction formulas are most
compactly expressed in terms of
$m_{\beta,\mathrm{norm}}^2=m_\beta^2/C_e^2$.

Using Eq.~\eqref{eq:first-row-moduli}, one obtains
\begin{equation}
m_{\beta,\mathrm{norm}}^2
=
\frac{m_\beta^2}{C_e^2}
=
c_{12}^2c_{13}^2m_1^2
+
s_{12}^2c_{13}^2m_2^2
+
s_{13}^2m_3^2.
\label{eq:mbeta}
\end{equation}
The three coefficients on the right-hand side obey
\begin{equation}
c_{12}^2c_{13}^2
+
s_{12}^2c_{13}^2
+
s_{13}^2
=
1,
\label{eq:first-row-normalization}
\end{equation}
because $c_{12}^2+s_{12}^2=1$ and
$c_{13}^2+s_{13}^2=1$.

We adopt the signed definitions
\begin{equation}
\Delta m_{21}^2
\equiv
m_2^2-m_1^2,
\qquad
\Delta m_{31}^2
\equiv
m_3^2-m_1^2,
\qquad
\Delta m_{32}^2
\equiv
m_3^2-m_2^2.
\label{eq:mass-squared-differences}
\end{equation}
These quantities satisfy
\begin{equation}
\Delta m_{31}^2
=
\Delta m_{32}^2+\Delta m_{21}^2.
\label{eq:mass-squared-difference-identity}
\end{equation}
Substituting
$m_2^2=m_1^2+\Delta m_{21}^2$ and
$m_3^2=m_1^2+\Delta m_{31}^2$ into
Eq.~\eqref{eq:mbeta}, and then using
Eq.~\eqref{eq:first-row-normalization}, gives
\begin{equation}
m_1^2
=
m_{\beta,\mathrm{norm}}^2
-
s_{12}^2c_{13}^2\Delta m_{21}^2
-
s_{13}^2\Delta m_{31}^2.
\label{eq:m1-corrected}
\end{equation}

Adding $\Delta m_{21}^2$ to
Eq.~\eqref{eq:m1-corrected} and using
$\Delta m_{32}^2=\Delta m_{31}^2-\Delta m_{21}^2$ yields
\begin{equation}
m_2^2
=
m_{\beta,\mathrm{norm}}^2
+
c_{12}^2c_{13}^2\Delta m_{21}^2
-
s_{13}^2\Delta m_{32}^2.
\label{eq:m2-corrected}
\end{equation}
Similarly, adding $\Delta m_{31}^2$ to
Eq.~\eqref{eq:m1-corrected} gives
\begin{equation}
m_3^2
=
m_{\beta,\mathrm{norm}}^2
+
c_{12}^2c_{13}^2\Delta m_{31}^2
+
s_{12}^2c_{13}^2\Delta m_{32}^2.
\label{eq:m3-corrected}
\end{equation}
Equations~\eqref{eq:m1-corrected}--\eqref{eq:m3-corrected}
are equivalent representations of the same weighted beta-decay
relation. They hold for either light-neutrino mass ordering, provided
the signed definitions in
Eq.~\eqref{eq:mass-squared-differences} are used consistently.

In the complete Euler-like parametrization employed in
Ref.~\cite{XingFSS}, the relevant first-row elements of $U$ and $R$
have simple exact forms. Before dividing by any first-row mixing
factor, taking the absolute values of the corresponding FSS
relations for the fixed pairing $(\nu_i,N_i)$ and column assignment
adopted in Ref.~\cite{XingFSS} gives
\begin{align}
m_1c_{12}^2c_{13}^2c_{14}^2c_{15}^2c_{16}^2
&=
M_1s_{14}^2c_{15}^2c_{16}^2,
\nonumber\\
m_2s_{12}^2c_{13}^2c_{14}^2c_{15}^2c_{16}^2
&=
M_2s_{15}^2c_{16}^2,
\nonumber\\
m_3s_{13}^2c_{14}^2c_{15}^2c_{16}^2
&=
M_3s_{16}^2.
\label{eq:first-row-FSS-undivided}
\end{align}
These undivided relations remain the appropriate starting point when
one or more of the first-row mixing factors vanish. When the factors
cancelled or divided out are nonzero and the Euler parametrization is
nonsingular, Eq.~\eqref{eq:first-row-FSS-undivided} gives
\begin{equation}
\frac{m_1}{M_1}
=
\frac{t_{14}^2}{c_{12}^2c_{13}^2},
\qquad
\frac{m_2}{M_2}
=
\frac{t_{15}^2}{s_{12}^2c_{13}^2c_{14}^2},
\qquad
\frac{m_3}{M_3}
=
\frac{t_{16}^2}{s_{13}^2c_{14}^2c_{15}^2}.
\label{eq:first-row-FSS}
\end{equation}
These equalities are exact, wherever they are well defined, within
the stated parametrization, FSS assignment, and paired column
ordering, and relate the physical heavy mass eigenvalues $M_i$ to
the first-row mixing parameters. They should be distinguished from
the exact relation between $Y_\nu$ and $R$ in
Eq.~\eqref{eq:R-Y-exact-components}, as well as from the relation
$\widehat{M}_i=M_i-m_i$ between the physical heavy eigenvalues and
the right-handed Majorana mass parameters. A different discrete
light--heavy pairing would require a corresponding permutation of the
mass labels, mixing columns, and active--heavy angles.

For positive masses, nonzero corresponding active--heavy mixing
angles, and nonzero factors appearing in the denominators of
Eq.~\eqref{eq:first-row-FSS}, that equation may equivalently be
solved as
\begin{equation}
M_1
=
m_1\frac{c_{12}^2c_{13}^2}{t_{14}^2},
\qquad
M_2
=
m_2\frac{s_{12}^2c_{13}^2c_{14}^2}{t_{15}^2},
\qquad
M_3
=
m_3\frac{s_{13}^2c_{14}^2c_{15}^2}{t_{16}^2}.
\label{eq:heavy-masses-from-FSS}
\end{equation}
Equation~\eqref{eq:first-row-FSS} requires the displayed denominator
factors to be nonzero and the Euler parametrization to be
nonsingular. The divided reconstruction formulas in
Eq.~\eqref{eq:heavy-masses-from-FSS} additionally require the
corresponding $t_{1j}$ to be nonzero. If a factor that was divided
out vanishes, the limiting case must instead be analyzed from the
undivided component relation
$m_i|U_{ei}|^2=M_i|R_{ei}|^2$, or equivalently from
Eq.~\eqref{eq:first-row-FSS-undivided}, rather than from
Eq.~\eqref{eq:first-row-FSS}. If only $t_{1j}=0$ while the
corresponding denominator in Eq.~\eqref{eq:first-row-FSS} is
nonzero, that equation implies $m_i=0$ for finite positive $M_i$.
The corresponding $M_i$ nevertheless cannot be reconstructed by
division by the vanishing angle.

Squaring Eq.~\eqref{eq:heavy-masses-from-FSS} and substituting
Eqs.~\eqref{eq:m1-corrected}--\eqref{eq:m3-corrected}, one obtains
\begin{equation}
M_1^2
=
\frac{c_{12}^4c_{13}^4}{t_{14}^4}
\left[
m_{\beta,\mathrm{norm}}^2
-
s_{12}^2c_{13}^2\Delta m_{21}^2
-
s_{13}^2\Delta m_{31}^2
\right].
\label{eq:M1-corrected}
\end{equation}
Similarly,
\begin{equation}
M_2^2
=
\frac{s_{12}^4c_{13}^4c_{14}^4}{t_{15}^4}
\left[
m_{\beta,\mathrm{norm}}^2
+
c_{12}^2c_{13}^2\Delta m_{21}^2
-
s_{13}^2\Delta m_{32}^2
\right],
\label{eq:M2-corrected}
\end{equation}
and
\begin{equation}
M_3^2
=
\frac{s_{13}^4c_{14}^4c_{15}^4}{t_{16}^4}
\left[
m_{\beta,\mathrm{norm}}^2
+
c_{12}^2c_{13}^2\Delta m_{31}^2
+
s_{12}^2c_{13}^2\Delta m_{32}^2
\right].
\label{eq:M3-corrected}
\end{equation}

The square brackets in
Eqs.~\eqref{eq:M1-corrected}--\eqref{eq:M3-corrected} are,
respectively, exactly $m_1^2$, $m_2^2$, and $m_3^2$. The prefactors
outside the brackets are the squares of the corresponding
light-to-heavy conversion factors in
Eq.~\eqref{eq:heavy-masses-from-FSS}. This observation provides an
immediate algebraic check of all three reconstructed heavy masses.
Once the physical heavy eigenvalues have been reconstructed, the
corresponding right-handed Majorana mass parameters follow, in the
canonical domain of
Eq.~\eqref{eq:canonical-domain-mass-reconstruction}, from
$\widehat{M}_i=M_i-m_i>0$. Outside that domain, the physical singlet
Takagi mass is instead $|M_i-m_i|$, with the corresponding phase
convention adjusted as described above.

These results differ from Eq.~(20) of Ref.~\cite{XingFSS} in several
essential respects. First, no common factor
$1/(s_{12}^2c_{13}^2+s_{13}^2)$ appears in the corrected formulas.
The coefficient multiplying the chosen reference mass squared is
unity because the three weights in Eq.~\eqref{eq:mbeta} satisfy the
normalization identity in
Eq.~\eqref{eq:first-row-normalization}. Second, the expression for
$M_2^2$ necessarily contains the contribution
$+c_{12}^2c_{13}^2\Delta m_{21}^2$. Third, the expression for
$M_3^2$ necessarily contains the contribution
$+c_{12}^2c_{13}^2\Delta m_{31}^2$. Omitting these terms makes the
quantities inside the square brackets unequal to the corresponding
light-neutrino masses squared and hence renders the resulting
heavy-mass reconstruction inconsistent with
Eq.~\eqref{eq:mbeta}.

The positivity of $M_i^2$ does not yield a universal mass-ordering
criterion of the form stated in Ref.~\cite{XingFSS}. Because the
square brackets in
Eqs.~\eqref{eq:M1-corrected}--\eqref{eq:M3-corrected} are identities
for the corresponding $m_i^2$, their nonnegativity contains no
independent information beyond the physical requirement
$m_i^2\geq0$. The prefactors are also nonnegative wherever the
reconstruction is well defined. Positivity consequently imposes no
new ordering relation.

The signs of $\Delta m_{31}^2$ and $\Delta m_{32}^2$ depend on the
light-neutrino mass ordering, so the terms containing these
quantities must be interpreted using the signed definitions in
Eq.~\eqref{eq:mass-squared-differences}. Moreover, the ordering of the
physical heavy neutrinos is not fixed by the light-neutrino ordering
alone, because each $M_i$ depends on a distinct active--heavy mixing
angle through $t_{14}$, $t_{15}$, or $t_{16}$. The indices $i$ label
the chosen FSS pairs and do not by themselves imply an ascending
ordering $M_1<M_2<M_3$; ordering the reconstructed heavy spectrum
may require a subsequent simultaneous permutation of the pair labels
and their associated mixing parameters. Only after these mixing
angles are independently specified or constrained can the physical
heavy spectrum be reconstructed. The corresponding right-handed
Majorana parameters
$\widehat{M}_i=M_i-m_i$ are then determined as well within the
canonical domain. Their ordering need not coincide exactly with that
of the physical heavy eigenvalues outside the conventional seesaw
hierarchy. The resulting heavy-mass determination is therefore
conditional on the chosen beta-decay convention and input, the
active--heavy mixing parameters, and the fixed FSS pairing, rather
than being a prediction based on the light-neutrino ordering alone.

\section{Further phenomenological implications}
\label{sec:phenomenology}

The exact FSS solution has several additional consequences that are
not captured by the separate elementwise relations alone. In this
section, we clarify its implications for neutrinoless double-beta
decay, the relation between active--heavy mixing and neutrino Yukawa
couplings, the number of independent mixing parameters, and the
connection between the light-neutrino ordering and the spectra of the
physical heavy eigenvalues and right-handed Majorana parameters. As
in the preceding sections, ``exact'' refers to the algebraic
tree-level FSS relations, with all running masses and couplings
appearing in a given relation evaluated at a common renormalization
scale. Whenever the right-handed Majorana parameters are interpreted
as positive physical singlet masses, we work in the canonical domain
$M_i>m_i$, for which
$\widehat{M}_i=M_i-m_i>0$.

\subsection{Neutrinoless double-beta decay}
\label{subsec:0nubb}

The $ee$ component of the exact seesaw relation
$UD_\nu U^T+RD_NR^T=\mathbf{0}_{3\times3}$ gives
\begin{equation}
\sum_{i=1}^{3}m_iU_{ei}^2
=
-\sum_{i=1}^{3}M_iR_{ei}^2.
\label{eq:ee-seesaw-sum-rule}
\end{equation}
This equation is an exact identity between entries of the tree-level
neutral-fermion mass matrix. It may be referred to as the exact
seesaw sum rule. The quantities $M_i$ in this identity are the exact
positive heavy Takagi eigenvalues of the complete broken-phase mass
matrix, rather than the right-handed Majorana parameters
$\widehat{M}_i=M_i-m_i$. The identity must not, however, be
interpreted as an equality between the physical light- and
heavy-neutrino exchange contributions to the neutrinoless
double-beta-decay amplitude
\cite{Rodejohann2011,FangLiZhangZhu2024}.

For a Majorana neutrino mass eigenstate of mass $m_n$, the
lepton-number-violating part of the propagator produces a schematic
contribution proportional to $m_n/(p^2-m_n^2)$, where $p$ denotes the
virtual neutrino four-momentum inside the nucleus and is
predominantly spacelike. At the level of the neutrino-line mass and
propagator dependence, the corresponding factor has the schematic
form
\begin{equation}
\mathcal{A}_{0\nu\beta\beta}^{\mathrm{prop}}
\propto
\sum_{i=1}^{3}
U_{ei}^2
\frac{m_i}{p^2-m_i^2}
+
\sum_{i=1}^{3}
R_{ei}^2
\frac{M_i}{p^2-M_i^2}.
\label{eq:0nubb-propagator-amplitude}
\end{equation}
Equation~\eqref{eq:0nubb-propagator-amplitude} displays only the
mass and propagator dependence of the exchanged-neutrino line. In a
complete nuclear calculation, the relevant weak-interaction
normalization, phase-space factor, hadronic currents, and nuclear
matrix element must also be included, and the nuclear matrix element
itself depends on the exchanged-neutrino mass
\cite{FangLiZhangZhu2024}.

For the three light neutrinos, $m_i^2\ll |p^2|$, and hence
\begin{equation}
\frac{m_i}{p^2-m_i^2}
=
\frac{m_i}{p^2}
\left[
1+\mathcal{O}\!\left(\frac{m_i^2}{p^2}\right)
\right]
\simeq
\frac{m_i}{p^2},
\qquad
m_i^2\ll |p^2|.
\label{eq:light-propagator-limit}
\end{equation}
The standard light-neutrino contribution is therefore controlled by
the complex effective Majorana mass
\begin{equation}
m_{ee}
\equiv
\sum_{i=1}^{3}m_iU_{ei}^2.
\label{eq:mee-definition}
\end{equation}
The experimentally relevant quantity associated with this
combination is $|m_{ee}|$.

By contrast, for heavy Majorana neutrinos satisfying
$M_i^2\gg |p^2|$,
\begin{equation}
\frac{M_i}{p^2-M_i^2}
=
-\frac{1}{M_i}
\left[
1+\mathcal{O}\!\left(\frac{p^2}{M_i^2}\right)
\right]
\simeq
-\frac{1}{M_i},
\qquad
M_i^2\gg |p^2|.
\label{eq:heavy-propagator-limit}
\end{equation}
Their exchange contribution is therefore controlled schematically at
the propagator level by
\begin{equation}
\mathcal{A}_{0\nu\beta\beta}^{(N),\mathrm{prop}}
\propto
-\sum_{i=1}^{3}\frac{R_{ei}^2}{M_i},
\label{eq:heavy-0nubb-amplitude}
\end{equation}
together with the appropriate short-range nuclear matrix element in
the complete physical amplitude. The heavy-neutrino contribution is
thus not governed by $\sum_iM_iR_{ei}^2$, despite the exact
mass-matrix identity in Eq.~\eqref{eq:ee-seesaw-sum-rule}.

Applying the exact FSS relation
\begin{equation}
R_{ei}
=
i\eta_i\sqrt{\frac{m_i}{M_i}}\,U_{ei}
\label{eq:FSS-electron-column-relation}
\end{equation}
to each term in Eq.~\eqref{eq:heavy-0nubb-amplitude} gives
\begin{equation}
-\frac{R_{ei}^2}{M_i}
=
\frac{m_iU_{ei}^2}{M_i^2},
\label{eq:FSS-heavy-0nubb-scaling}
\end{equation}
where $\eta_i^2=1$ has been used. For a nonvanishing corresponding
light-neutrino term, the heavy-neutrino term associated with the same
light--heavy pair is therefore suppressed at the propagator level
relative to that light-neutrino term by a factor of order
$|p^2|/M_i^2$ in the regime $M_i^2\gg |p^2|$. If
$m_iU_{ei}^2=0$, both terms in this pairwise comparison vanish and
their ratio is not defined. This comparison does not by itself
include the difference between the long-range and short-range
nuclear matrix elements. It therefore establishes the propagator
scaling but should not be interpreted as a quantitative prediction
for the ratio of the complete nuclear decay amplitudes.

Taking the absolute value of
Eq.~\eqref{eq:ee-seesaw-sum-rule} gives
\begin{equation}
|m_{ee}|
=
\left|
\sum_{i=1}^{3}M_iR_{ei}^2
\right|.
\label{eq:mee-heavy-sum-rule}
\end{equation}
Equation~\eqref{eq:mee-heavy-sum-rule} is a mass-matrix sum rule, not
an equality between the two observable exchange amplitudes. The
physical light- and heavy-neutrino contributions contain different
momentum dependences and, in general, different nuclear matrix
elements. In particular, the exact seesaw sum rule does not imply a
cancellation of the physical light- and heavy-exchange amplitudes in
the light- and heavy-mass regimes described above.

More generally, if an additional neutrino has a mass comparable to
the characteristic nuclear momentum, neither the light-mass
expansion in Eq.~\eqref{eq:light-propagator-limit} nor the
heavy-mass expansion in Eq.~\eqref{eq:heavy-propagator-limit} is
valid. In that intermediate-mass regime, the full mass-dependent
propagator and nuclear matrix element must be retained
\cite{FangLiZhangZhu2024}.

\subsection{Active--heavy mixing and neutrino Yukawa couplings}
\label{subsec:yukawa-scaling}

In the canonical domain $M_i>m_i$, choose the sterile weak basis in
which the right-handed Majorana mass matrix is positive and diagonal,
\begin{equation}
M_{\mathrm R}
=
\operatorname{diag}
\left(
\widehat{M}_1,\widehat{M}_2,\widehat{M}_3
\right),
\qquad
\widehat{M}_i
=
M_i-m_i>0.
\label{eq:MR-positive-diagonal-phenomenology}
\end{equation}
At the common scale at which the exact FSS alignment is imposed, the
exact unitary completion derived in
Sec.~\ref{sec:exact-structure} then gives
\begin{equation}
R_{\alpha i}
=
\frac{v(Y_\nu)_{\alpha i}}
{\sqrt{M_i(M_i+m_i)}},
\label{eq:R-Y-phenomenology}
\end{equation}
where we use the convention
\begin{equation}
M_D
=
vY_\nu,
\qquad
v
=
\langle\phi^0\rangle
\simeq
174~\mathrm{GeV}.
\label{eq:Dirac-mass-convention}
\end{equation}
Here $M_i$ is the exact positive heavy-neutrino mass eigenvalue of the
complete broken-phase neutral-lepton mass matrix. The corresponding
positive right-handed Majorana mass is
$\widehat{M}_i=M_i-m_i$, and the $i$th column of $Y_\nu$ couples to
the paired singlet state $\mathcal{N}_i$ introduced in
Sec.~\ref{sec:leptogenesis}. Thus the active--heavy mixing matrix
$R$ refers to the exact broken-phase mass eigenstates $N_i$, whereas
$Y_\nu$ and $\widehat{M}_i$ describe the singlet fields
$\mathcal{N}_i$ in the unbroken-phase formulation.

Combining Eq.~\eqref{eq:R-Y-phenomenology} with the exact FSS column
relation gives
\begin{equation}
(Y_\nu)_{\alpha i}
=
i\eta_i
\frac{\sqrt{m_i(M_i+m_i)}}{v}\,
U_{\alpha i}.
\label{eq:FSS-yukawa-scaling}
\end{equation}
Equivalently, using
$U_{\alpha i}=V_{\alpha i}/\sqrt{1+r_i}$, one has
\begin{equation}
R_{\alpha i}
=
i\eta_i
\sqrt{\frac{m_i}{M_i+m_i}}\,
V_{\alpha i},
\qquad
(Y_\nu)_{\alpha i}
=
i\eta_i
\frac{\sqrt{m_iM_i}}{v}\,
V_{\alpha i}.
\label{eq:mixing-yukawa-opposite-scaling}
\end{equation}

Equation~\eqref{eq:mixing-yukawa-opposite-scaling} demonstrates that
active--heavy mixing and Yukawa strength scale differently with the
heavy-neutrino mass. For fixed $m_i$, $V_{\alpha i}$, and $v$, the
mixing $|R_{\alpha i}|$ decreases as
$(M_i+m_i)^{-1/2}$, approaching the familiar $M_i^{-1/2}$ behavior
when $M_i\gg m_i$, whereas the Yukawa coupling
$|(Y_\nu)_{\alpha i}|$ increases as $M_i^{1/2}$. Consequently, a
very small active--heavy mixing angle does not imply a very weak
neutrino Yukawa interaction. In a high-scale seesaw, the
active--heavy mixing can be extremely suppressed while the
corresponding Yukawa coupling remains sizable and may eventually
leave the perturbative domain.

To express the seesaw relation quantitatively, define the norm of the
$i$th Yukawa column by
\begin{equation}
y_i^2
\equiv
(Y_\nu^\dagger Y_\nu)_{ii}
=
\sum_{\alpha}
|(Y_\nu)_{\alpha i}|^2.
\label{eq:yi-definition}
\end{equation}
Using Eq.~\eqref{eq:FSS-yukawa-scaling}, one finds
\begin{align}
y_i^2
&=
\frac{m_i(M_i+m_i)}{v^2}
\sum_{\alpha}|U_{\alpha i}|^2
\nonumber\\
&=
\frac{m_iM_i}{v^2},
\label{eq:yi-FSS-relation}
\end{align}
where the second line uses the exact FSS identity
\begin{equation}
\sum_{\alpha}|U_{\alpha i}|^2
=
(U^\dagger U)_{ii}
=
\frac{1}{1+r_i}
=
\frac{M_i}{M_i+m_i}.
\label{eq:U-column-norm-expansion}
\end{equation}
The corresponding relation between the physical light and heavy mass
eigenvalues and the Yukawa-column norm is therefore exact:
\begin{equation}
m_i
=
\frac{v^2y_i^2}{M_i}.
\label{eq:usual-seesaw-scaling}
\end{equation}
In terms of the positive right-handed Majorana mass
$\widehat{M}_i=M_i-m_i$, the same identity may be written as
\begin{equation}
y_i^2
=
\frac{m_i(\widehat{M}_i+m_i)}{v^2}.
\label{eq:yi-Mhat-relation}
\end{equation}
In the conventional seesaw hierarchy
$m_i\ll\widehat{M}_i$, this reduces to the familiar approximate
relation
$m_i\simeq v^2y_i^2/\widehat{M}_i$.

The smallness of the light-neutrino masses therefore need not be
attributed solely to weak Yukawa interactions. It may result from a
large right-handed Majorana mass scale, small Yukawa couplings, or a
combination of the two. These exact algebraic relations hold at the
scale at which the FSS alignment is imposed. If FSS is instead
specified as a boundary condition at a different scale,
renormalization-group evolution and threshold corrections can modify
the alignment and the corresponding Yukawa-column relations
\cite{CooperKingLuhn2012}.

\subsection{Parameter counting and phase structure}
\label{subsec:parameter-counting}

For a fixed light--heavy pairing and chosen nondegenerate Takagi mass
bases, the exact FSS solution for the upper charged-current mixing
blocks derived in Sec.~\ref{sec:exact-structure} can be written as
\begin{equation}
U
=
V\operatorname{diag}\!\left(
\frac{1}{\sqrt{1+r_1}},
\frac{1}{\sqrt{1+r_2}},
\frac{1}{\sqrt{1+r_3}}
\right),
\label{eq:U-parameter-counting}
\end{equation}
and
\begin{equation}
R
=
iV\operatorname{diag}\!\left(
\eta_1\sqrt{\frac{r_1}{1+r_1}},
\eta_2\sqrt{\frac{r_2}{1+r_2}},
\eta_3\sqrt{\frac{r_3}{1+r_3}}
\right),
\label{eq:R-parameter-counting}
\end{equation}
where $V$ is unitary, $r_i=m_i/M_i$, and $\eta_i=\pm1$. Thus $U$
and $R$ do not contain two independent sets of column directions.
They share the same unitary matrix $V$, and each heavy-neutrino
column differs from the corresponding light-neutrino column only by
a normalization factor and the discrete relative factor $i\eta_i$.

A general $3\times3$ unitary matrix contains three mixing angles and
six phases. Three row phases may be removed by rephasing the charged
lepton fields. For three nondegenerate and nonzero Majorana
neutrinos, the remaining continuous physical mixing information may
therefore be parametrized by three mixing angles and three
CP-violating phases, conventionally identified as one Dirac-type
phase and two Majorana-type phases. If one light-neutrino mass
vanishes, one Majorana phase becomes unphysical. Degeneracies may
reduce the number of physical mixing parameters further, as
discussed below. The FSS mixing structure additionally contains the
three nonnegative ratios $r_i$. The factors $\eta_i=\pm1$ can be
changed by allowed Majorana column-sign redefinitions and therefore
do not constitute additional physical parameters or continuous
sources of CP violation.

This parameter count refers to the physical mixing information
contained in the aligned $U$ and $R$ blocks for a fixed paired column
assignment. In addition to this mixing information, a complete
specification of the spectrum contains the three nonnegative
light-neutrino masses $m_i$ and the three positive heavy-neutrino
mass eigenvalues $M_i$, subject to the canonical inequalities
$M_i>m_i$ when a positive singlet-mass interpretation is required and
to whatever further relations or inputs are imposed in a particular
phenomenological analysis. Since $r_i=m_i/M_i$, the ratios $r_i$ are
not independent of both sets of masses if the latter are counted
separately. The corresponding positive right-handed Majorana masses
are then fixed by $\widehat{M}_i=M_i-m_i$.

Thus, for nonzero and nondegenerate light and heavy spectra and a
fixed light--heavy pairing, the FSS neutrino sector contains twelve
continuous physical parameters: six mass eigenvalues, three mixing
angles, and three CP-violating phases. This count excludes the
charged-lepton masses. The discrete factors $\eta_i$, the discrete
choice of light--heavy pairing, and the sterile weak-basis matrix
$W$ do not add physical continuous parameters.

The arbitrary unitary matrix $W$ appearing in the full unitary
completion of Sec.~\ref{sec:exact-structure} represents a sterile
weak-basis choice and does not introduce additional physical
charged-current mixing parameters in the canonical type-I framework.
It follows that the active--heavy angles and phases appearing in a
general Euler-like parametrization of the complete $6\times6$ mixing
matrix cannot remain independent after the exact FSS conditions are
imposed. They must satisfy the nonlinear constraints required to
reproduce Eqs.~\eqref{eq:U-parameter-counting} and
\eqref{eq:R-parameter-counting}. In particular, the continuous phases
appearing in individual elements of $R$ are inherited from the same
unitary matrix $V$ that determines $U$. They are not independent
high-energy phases that can be varied freely while the light-neutrino
mixing matrix is held fixed.

In the conventional seesaw hierarchy, the same reduction can be
expressed in the Casas--Ibarra language. With the convention
\begin{equation}
M_D
=
iU_\nu D_\nu^{1/2}
\mathbb{O}
D_{\widehat{N}}^{1/2},
\qquad
D_{\widehat{N}}
\equiv
\operatorname{diag}
\left(
\widehat{M}_1,\widehat{M}_2,\widehat{M}_3
\right),
\label{eq:Casas-Ibarra-phenomenology}
\end{equation}
where $\mathbb{O}$ is complex orthogonal, the limits
$U_\nu\simeq V$ and
$D_{\widehat{N}}\simeq D_N$ give
\begin{equation}
\mathbb{O}
\simeq
E
\equiv
\operatorname{diag}(\eta_1,\eta_2,\eta_3),
\label{eq:Casas-Ibarra-FSS}
\end{equation}
up to signed permutations associated with the chosen pairing and
column conventions. Thus exact FSS approaches the form-dominant
limit in which the Casas--Ibarra matrix contains no independent
continuous complex angles
\cite{CasasIbarra2001,King2009,ChoubeyKingMitra2010}. This
Casas--Ibarra identification is a conventional-seesaw-limit
statement: the exact completion distinguishes
$U$ from $V$ and the broken-phase heavy eigenvalues $M_i$ from the
singlet masses $\widehat{M}_i$.

This reduction of parameter freedom is also reflected in the exact
orthogonality relations
\begin{equation}
(U^\dagger U)_{ik}
=
(R^\dagger R)_{ik}
=
0,
\qquad
i\neq k.
\label{eq:shared-column-orthogonality}
\end{equation}
Exact FSS also implies
$U^\dagger U+R^\dagger R=\mathbf{1}$, as derived in
Eq.~\eqref{eq:column-partial-unitarity}. This relation is not a
generic block-unitarity condition of the complete $6\times6$ mixing
matrix, but a special consequence of exact FSS alignment and the
fixed one-to-one light--heavy pairing. Accordingly, a
phenomenological analysis based on a general Euler-like
parametrization must impose the complete FSS constraints before
treating its active--heavy angles and phases as independent
phenomenological parameters. Exact parametrizations of seesaw models
likewise show that the light- and heavy-sector masses and mixings are
correlated rather than independent
\cite{BlennowFernandezMartinez2011,ChenHuZhou2025}.

An exact degeneracy among nonzero light eigenvalues $m_i$ introduces
a real orthogonal basis freedom within the corresponding light
Majorana subspace. If several light eigenvalues vanish, the massless
subspace possesses a larger unitary basis freedom. An exact
degeneracy among the positive physical heavy eigenvalues $M_i$
similarly introduces a real orthogonal basis freedom in the heavy
columns of the full mixing matrix. In either case, the individual
column directions and a family-separated pairing are meaningful only
after a definite mass basis within the degenerate subspace has been
specified.

Separately, an exact degeneracy among the positive right-handed
Majorana masses $\widehat{M}_i$ introduces a residual real orthogonal
freedom in the singlet mass basis relevant to the Yukawa couplings
and leptogenesis. The paired basis obtained by choosing
$W=\mathbf{1}$ provides a simultaneous representation in which
$M_{\mathrm R}$ and $Y_\nu^\dagger Y_\nu$ are diagonal. Nevertheless,
individual singlet states within an exactly degenerate subspace are
not basis-invariant objects, and an orthogonal rotation can change
the matrix representation of $Y_\nu^\dagger Y_\nu$ in that
subspace. Their decay and oscillation properties therefore require a
basis-invariant coherent treatment rather than the assignment of
independent nonresonant decay asymmetries to arbitrarily chosen
degenerate states. Since
$\widehat{M}_i=M_i-m_i$, degeneracy among the physical heavy
eigenvalues $M_i$ and degeneracy among the singlet masses
$\widehat{M}_i$ need not coincide outside the conventional seesaw
hierarchy.

\subsection{Light and heavy mass orderings}
\label{subsec:mass-orderings}

For the fixed paired column assignment used in
Ref.~\cite{XingFSS}, positive masses, nonzero corresponding
active--heavy mixing angles, and nonzero factors appearing in the
relevant denominators, the first-row FSS relations imply
\begin{equation}
M_1
=
m_1
\frac{c_{12}^2c_{13}^2}{t_{14}^2},
\qquad
M_2
=
m_2
\frac{s_{12}^2c_{13}^2c_{14}^2}{t_{15}^2},
\qquad
M_3
=
m_3
\frac{s_{13}^2c_{14}^2c_{15}^2}{t_{16}^2}.
\label{eq:heavy-masses-linear}
\end{equation}
The divided expressions in
Eq.~\eqref{eq:heavy-masses-linear} apply only when the factors divided
out in their derivation are nonzero and the Euler parametrization is
nonsingular. Zero-angle, massless, vanishing-first-row-element, or
singular-parametrization limits must instead be analyzed from the
undivided relations in
Eq.~\eqref{eq:first-row-FSS-undivided}, or directly from
$m_i|U_{ei}|^2=M_i|R_{ei}|^2$. If only a corresponding
active--heavy mixing angle vanishes while the other factors entering
the relevant relation are nonzero, the undivided FSS relation implies
$m_i=0$ for finite positive $M_i$. In that limit, the divided
expression cannot be used to reconstruct the corresponding $M_i$.
These equations show that the physical heavy-neutrino spectrum
depends not only on the light-neutrino masses but also on three
distinct active--heavy mixing angles.

The following ratios apply only when the masses and mixing factors
appearing in their denominators are nonzero. Massless, zero-angle,
vanishing-first-row-element, or singular-parametrization limits must
instead be analyzed from
Eq.~\eqref{eq:first-row-FSS-undivided}, or directly from the
undivided component relation
$m_i|U_{ei}|^2=M_i|R_{ei}|^2$. The ratio of the first two physical
heavy masses is
\begin{equation}
\frac{M_2}{M_1}
=
\frac{m_2}{m_1}
\frac{s_{12}^2}{c_{12}^2}
c_{14}^2
\frac{t_{14}^2}{t_{15}^2},
\label{eq:M2-M1-ratio}
\end{equation}
whereas
\begin{equation}
\frac{M_3}{M_1}
=
\frac{m_3}{m_1}
\frac{s_{13}^2c_{14}^2c_{15}^2}
{c_{12}^2c_{13}^2}
\frac{t_{14}^2}{t_{16}^2}.
\label{eq:M3-M1-ratio}
\end{equation}
Even after the light-neutrino ordering and masses are fixed, the
independent ratios $t_{14}/t_{15}$ and $t_{14}/t_{16}$ can change or
reverse the ordering of the physical heavy states. The
light-neutrino mass ordering therefore does not determine the
physical heavy-neutrino mass ordering by itself.

The corrected reconstruction formulas in
Eqs.~\eqref{eq:M1-corrected}--\eqref{eq:M3-corrected} determine the
physical heavy mass eigenvalues only after the convention-dependent
normalized beta-decay parameter $m_{\beta,\mathrm{norm}}$ and the
three relevant active--heavy mixing angles have also been specified.
Using an experimentally quoted beta-decay parameter in this
reconstruction requires that its spectral-normalization and
nonunitarity conventions be matched to those of
Eq.~\eqref{eq:normalized-mbeta-definition}. In the canonical domain,
the corresponding positive right-handed Majorana masses then follow
from
$\widehat{M}_i=M_i-m_i$. Outside the conventional hierarchy
$m_i\ll M_i$, but still within the canonical domain $M_i>m_i$, the
ordering of the $\widehat{M}_i$ need not coincide identically with
the ordering of the physical heavy eigenvalues $M_i$. If the
algebraic construction is formally extended outside the canonical
domain, the corresponding physical singlet Takagi values are
$|M_i-m_i|$ rather than $M_i-m_i$. The reconstruction is therefore
conditional rather than a direct prediction of either heavy ordering
from the light ordering.

There is, in addition, a labeling ambiguity. The FSS ansatz pairs the
light mass eigenstate $\nu_i$ with the exact heavy mass eigenstate
$N_i$ and, in the positive diagonal sterile basis of
Sec.~\ref{sec:exact-structure}, with the corresponding singlet column
associated with $\mathcal{N}_i$. The labels $i=1,2,3$ identify these
chosen pairs and do not by themselves imply an ascending ordering of
the physical heavy masses. The exact seesaw equation by itself does
not select a unique pairing between the light and physical heavy
spectra. A simultaneous relabeling of the heavy eigenstates and the
associated columns of $R$ leaves the physical theory unchanged,
while choosing a different light--heavy pairing corresponds to a
different discrete FSS assignment unless an additional flavor
symmetry or dynamical principle specifies the pairing. Any proposed
correlation between the two spectra must therefore state explicitly
how the light and heavy mass eigenstates are paired.

If the nonzero light spectrum or the physical heavy spectrum contains
an exact degeneracy, the corresponding real orthogonal basis freedom
must be fixed before an individual light--heavy pairing can be
assigned. A massless light subspace may possess a larger unitary
freedom. Separately, a degeneracy among the positive
$\widehat{M}_i$ makes the singlet basis nonunique and must be treated
independently when discussing the ordering or decay properties of the
right-handed Majorana states. In particular, individual decay
properties assigned to exactly degenerate singlet states are
basis-dependent unless formulated through an appropriate
basis-invariant coherent treatment.

The robust conclusion is that the FSS ansatz correlates each chosen
light--heavy pair through the ratio $m_i/M_i$ and the corresponding
mixing-column normalization. Without further assumptions, it does
not predict a universal correspondence between the global ordering
of the three light masses and that of either the three physical heavy
mass eigenvalues or, within the canonical domain, the corresponding
positive right-handed Majorana masses.

\section{Conclusions}
\label{sec:conclusions}

We have examined the family-separated seesaw ansatz of
Ref.~\cite{XingFSS}, in which, for a fixed light--heavy pairing and
paired column assignment, each light and heavy Majorana mass
eigenstate is required to cancel separately in the exact seesaw
relation. Although mathematically consistent, this ansatz defines a
highly constrained special realization of the canonical type-I
seesaw. Throughout our analysis, ``exact'' refers to the algebraic
tree-level mass-matrix relations and associated couplings evaluated
consistently at a common renormalization scale.

Imposing the FSS condition together with exact partial unitarity
forces the corresponding columns of the upper light- and
heavy-neutrino charged-current mixing blocks to be aligned. The two
blocks are determined by a common unitary matrix $V$ and the three
nonnegative ratios $r_i=m_i/M_i$, and both $U^\dagger U$ and
$R^\dagger R$ are exactly diagonal. They also satisfy the additional
exact FSS identity
\begin{equation*}
U^\dagger U+R^\dagger R
=
\mathbf{1}.
\end{equation*}
This relation is not a generic block-unitarity condition of the
complete $6\times6$ mixing matrix, but a special consequence of exact
FSS alignment and the fixed one-to-one light--heavy pairing. Thus
columns carrying distinct paired indices are mutually orthogonal,
independently of the CP-violating phases contained in $V$.
Alternative light--heavy pairings correspond to discrete alternative
FSS assignments rather than additional continuous parameters.

We have also constructed the most general unitary completion of the
fixed upper blocks and paired column assignment, up to sterile
weak-basis freedom. In the canonical domain
\begin{equation*}
M_i>m_i,
\end{equation*}
a sterile basis may be chosen in which the right-handed Majorana mass
matrix is positive and diagonal. In that basis, the reconstruction
gives
$M_D=iVE(D_\nu D_N)^{1/2}$ and
$M_{\mathrm R}=D_N-D_\nu$, with
$E=\operatorname{diag}(\eta_1,\eta_2,\eta_3)$. The quantities $M_i$
are the exact positive heavy Takagi eigenvalues of the complete
broken-phase neutral-lepton mass matrix, whereas the corresponding
positive singlet Majorana masses are
\begin{equation*}
\widehat{M}_i
=
M_i-m_i
>
0.
\end{equation*}
If the algebraic construction is formally continued to $M_i<m_i$,
the corresponding physical singlet Takagi value is
$|M_i-m_i|$ after an appropriate field rephasing; at $M_i=m_i$, the
singlet Majorana mass parameter vanishes. These formal domains do not
have the conventional heavy-singlet interpretation used in standard
seesaw leptogenesis.

The neutrino Yukawa matrix satisfies, at the scale of exact FSS
alignment, the exact identity
\begin{equation*}
Y_\nu^\dagger Y_\nu
=
\frac{1}{v^2}D_\nu D_N
=
\frac{1}{v^2}
\operatorname{diag}(m_1M_1,m_2M_2,m_3M_3),
\end{equation*}
so its columns associated with distinct paired indices are exactly
orthogonal in the paired basis where $M_{\mathrm R}$ is diagonal.
Equivalently, the reconstructed neutrino mass matrix is weak-basis
equivalent to three decoupled one-generation seesaw systems whose
charged-current flavor directions are supplied by $V$.

This result also makes explicit the distinction between the exact
broken-phase heavy states $N_i$, with masses $M_i$ and
charged-current mixing matrix $R$, and the unbroken-phase singlet
states $\mathcal{N}_i$, with masses $\widehat{M}_i$ and Yukawa
couplings $Y_\nu$. Standard singlet-neutrino decay calculations are
formulated in terms of $Y_\nu$ and $\widehat{M}_i$. The familiar
relations
$R_{\alpha i}\simeq v(Y_\nu)_{\alpha i}/M_i$ and
$\widehat{M}_i\simeq M_i$ are only leading-order approximations in
the conventional hierarchy. The exact completion instead gives
\begin{equation*}
R_{\alpha i}
=
\frac{v(Y_\nu)_{\alpha i}}
{\sqrt{M_i(M_i+m_i)}}.
\end{equation*}
Consequently, a decay-asymmetry expression written directly in terms
of the broken-phase quantities $R$ and $M_i$ must use this exact
conversion if it is claimed to apply beyond the conventional seesaw
limit.

The exact Yukawa-column orthogonality eliminates the CP-odd
interference terms entering the standard vacuum, nonresonant
one-loop unflavored and flavored decay asymmetries of the singlet
states $\mathcal{N}_i$. For every nondegenerate state with a nonzero
Yukawa decay width, the corresponding normalized decay asymmetries
vanish. If $m_i=0$, the paired Yukawa column and decay width vanish
identically, so no decay-produced lepton asymmetry is generated,
although the normalized asymmetry is then undefined. The same
cancellation occurs in the flavor-summed combination of rephasing
invariants used in Ref.~\cite{XingFSS}, even though individual
CP-odd invariants may remain nonzero. Low-energy leptonic CP
violation therefore does not imply a nonzero standard nonresonant
one-loop singlet-neutrino decay asymmetry when exact FSS alignment is
imposed at the decay scale.

This conclusion is deliberately restricted to the standard
nondegenerate perturbative one-loop decay source considered in
Ref.~\cite{XingFSS}. Exact or sufficiently close degeneracies among
the positive $\widehat{M}_i$ require a separate basis-invariant
resonant or flavor-coherent treatment and are not described by the
nonresonant formulas used here
\cite{PilaftsisUnderwood2004,DevMillingtonPilaftsisTeresi2014}.
Likewise, if FSS is imposed at a scale different from the decay
scale, renormalization-group evolution or threshold effects may
perturb the orthogonal Yukawa structure
\cite{CooperKingLuhn2012}. Finite-temperature, higher-loop, or other
nonstandard sources are also outside the no-go statement established
here. A recent preprint provides an example of a thermal two-loop
source based on lepton-flavor coherences
\cite{LiPilaftsis2026}. Such possibilities do not restore the
proposed result within the exact common-scale, standard nonresonant
one-loop framework, but they may lead to different phenomenology
once its assumptions are relaxed.

We have further corrected the reconstruction of the physical heavy
eigenvalues $M_i$ from the beta-decay input, the light-neutrino
mass-squared differences, and the active--heavy mixing angles. In the
nonunitary light sector, the unnormalized coefficient
\begin{equation*}
m_\beta^2
=
\sum_i m_i^2|U_{ei}|^2
\end{equation*}
must be distinguished, under the stated unresolved-spectrum fitting
convention, from the normalized weighted endpoint-shape parameter
\begin{equation*}
m_{\beta,\mathrm{norm}}^2
=
\frac{m_\beta^2}{C_e^2}.
\end{equation*}
The latter is a convention-dependent endpoint-shape parameter, not a
universal replacement for a complete beta-decay spectral analysis.
An experimental result obtained under an exactly unitary
three-neutrino hypothesis therefore cannot in general be inserted
directly into either definition without matching the treatment of
spectral normalization and possible nonunitarity.

The corrected heavy-mass formulas follow directly from the normalized
weighted mass relation and contain neither the additional common
denominator appearing in Ref.~\cite{XingFSS} nor the omissions in its
expressions for $M_2^2$ and $M_3^2$. In each corrected expression,
the bracketed quantity is exactly the corresponding $m_i^2$.
Positivity therefore provides no independent ordering criterion.
Once the physical eigenvalues $M_i$ are reconstructed, the positive
singlet Majorana masses follow in the canonical domain from
$\widehat{M}_i=M_i-m_i$. Outside that domain, the corresponding
Takagi value is $|M_i-m_i|$ rather than $M_i-m_i$.

The exact $ee$ seesaw identity is likewise a mass-matrix sum rule,
not an equality between observable light- and heavy-neutrino
exchange amplitudes in neutrinoless double-beta decay. The two
contributions contain different propagator dependences and generally
different nuclear matrix elements. In the regime
$M_i^2\gg|p^2|$, the physical heavy-neutrino exchange contribution is
controlled schematically at the propagator level by
$-\sum_iR_{ei}^2/M_i$, rather than by
$\sum_iM_iR_{ei}^2$. Under exact FSS, the heavy contribution from an
individual nonvanishing pair is suppressed relative to its light
counterpart at the propagator level by a factor of order
$|p^2|/M_i^2$. This scaling does not by itself include the different
long- and short-range nuclear matrix elements and therefore is not a
complete prediction for the ratio of observable decay amplitudes.

The exact FSS relations also show that small active--heavy mixing
does not by itself imply weak neutrino Yukawa couplings. The
Yukawa-column norm obeys
\begin{equation*}
y_i^2
\equiv
(Y_\nu^\dagger Y_\nu)_{ii}
=
\frac{m_iM_i}{v^2},
\end{equation*}
or equivalently $m_i=v^2y_i^2/M_i$. In terms of the positive
right-handed Majorana mass in the canonical domain,
$y_i^2=m_i(\widehat{M}_i+m_i)/v^2$, which reduces to the familiar
approximate seesaw relation
$m_i\simeq v^2y_i^2/\widehat{M}_i$ when
$m_i\ll\widehat{M}_i$. The light-neutrino mass scale may therefore
reflect a large singlet Majorana scale, small Yukawa couplings, or
both. Conversely, increasing the heavy scale at fixed $m_i$
increases the corresponding Yukawa-column norm even while the
active--heavy mixing decreases, eventually requiring attention to
perturbativity.

For a fixed light--heavy pairing and nonzero, nondegenerate light and
heavy spectra, the exact FSS neutrino sector contains twelve
continuous physical parameters: six mass eigenvalues, three mixing
angles, and three CP-violating phases. The matrices $U$ and $R$ share
the same continuous mixing information, while the factors
$\eta_i=\pm1$ are removable Majorana column-sign conventions and the
unitary matrix $W$ represents sterile weak-basis freedom. The choice
of light--heavy pairing is discrete and is not included as a
continuous parameter. A general Euler-like parametrization must
therefore satisfy nonlinear constraints before its active--heavy
angles and phases can be regarded as independent.

In the conventional seesaw hierarchy, this reduction is the
form-dominant limit of the Casas--Ibarra parametrization. Up to signed
permutations and the chosen pairing convention, the complex
orthogonal Casas--Ibarra matrix approaches
\begin{equation*}
\mathbb{O}
\simeq
E
=
\operatorname{diag}(\eta_1,\eta_2,\eta_3),
\end{equation*}
and contains no independent continuous complex angles. This
identification is a conventional-limit statement because the exact
construction distinguishes $U$ from $V$ and $M_i$ from
$\widehat{M}_i$.

Finally, the light-neutrino mass ordering does not determine either
the ordering of the physical heavy eigenvalues $M_i$ or, in the
canonical domain, that of the positive singlet Majorana masses
$\widehat{M}_i$. Each reconstructed $M_i$ depends on a distinct
active--heavy mixing angle, and the ordering of
$\widehat{M}_i=M_i-m_i$ need not coincide exactly with that of the
$M_i$ outside the conventional seesaw hierarchy. The result also
depends on the chosen light--heavy pairing. The indices $i$ label
the selected FSS pairs and do not by themselves imply an ascending
ordering of the heavy spectrum.

Degeneracies among the nonzero light masses, the physical heavy
eigenvalues, or the positive singlet Majorana masses introduce
distinct residual real orthogonal basis freedoms and must be treated
separately before assigning individual family pairings. A massless
light-neutrino subspace may possess a larger unitary freedom.
Moreover, individual decay properties assigned to exactly degenerate
singlet states are not basis invariant and require an appropriate
coherent treatment.

In summary, the family-separated ansatz enforces exact light--heavy
column alignment, admits a complete unitary reconstruction for a
fixed paired assignment, and fixes an exactly orthogonal Yukawa
structure at the scale where the alignment is imposed. Its
consequences eliminate the proposed correlation between low-energy
CP violation and the standard nonresonant one-loop decay asymmetries,
correct the heavy-mass reconstruction, and substantially restrict the
associated phenomenological parameter freedom and mass-ordering
interpretations. Radiative misalignment, resonant or coherent
dynamics, finite-temperature or higher-loop sources, and additional
interactions may lead to different phenomenology, but those
possibilities are not the standard exact common-scale nonresonant
FSS mechanism analyzed in Ref.~\cite{XingFSS}.

\bibliographystyle{apsrev4-2}
\bibliography{fss_comment}

\end{document}